\documentclass[traditabstract]{aa}
\usepackage{graphicx}
\usepackage{amsmath}
\usepackage{amssymb}
\usepackage{natbib}
\usepackage{txfonts}
\bibpunct{(}{)}{;}{a}{}{,} 
\newcommand{\vc}[1]{\textbf{\em #1}}

\begin{document}

\title{Reconnection in weakly stochastic B-fields in 2D}  
\author{K. Kulpa-Dybe{\l}\inst{1}\and
G. Kowal\inst{1}$^,$\inst{2}\and
K. Otmianowska-Mazur\inst{1}\and
A. Lazarian\inst{2}\and
E. Vishniac\inst{3}
}
\institute{Astronomical Observatory, Jagiellonian
University, ul Orla 171, 30-244 Krak\'ow, POLAND
\and
Department of Astronomy, University of Wisconsin,
475 North Charter Street, Madison, WI 53706, USA
\and
Department of Physics and Astronomy, McMaster University, 1280 Main Street West, Hamilton, ON L8S 4M1, CANADA}
\authorrunning{K. Kulpa-Dybel et al.}
\titlerunning{}
\offprints{K. Kulpa-Dybel}
\mail{kulpa@oa.uj.edu.pl}
\date{}
\abstract{We study two dimensional turbulent magnetic reconnection in a compressible fluid
in the gas pressure dominated limit. We use open boundary conditions and start from a Harris current sheet configuration 
with a uniform total pressure.  A small perturbation to the vector potential initiates 
laminar reconnection at the Sweet-Parker rate, which is allowed to evolve for several 
dynamical times. Subsequently sub-Alfvenic turbulence is produced through random 
forcing at small wave numbers. The magnetic field topology near the current sheet is strongly
affected by the turbulence.  However, we find that the resulting reconnection
speed depends on the resistivity.  In contrast to previous results in three
dimensions, we find no evidence for fast reconnection.  The reconnection speed
exhibits large variations but the time averages increase smoothly with the
strength of the turbulence.}

\maketitle

\section{Introduction}
Radio observations of a wide variety of astrophysical bodies (the Sun, spiral
galaxies, the Earth) show that a magnetic field is usually present
\citep[see][for review]{pri-00}.  These magnetic fields have significant large
scale components, i.e on the scale of the objects themselves.  The origin of
these fields is typically ascribed to the operation of a large scale
magnetohydrodynamic dynamo \citep{ruz-88,par-79,par-92,han-04,laz-08}.  Since
dynamo theory involves the twisting and folding of field lines it is important
that there is some process which can lead to efficient smoothing of the small
scale components of the field.  In other words, we need to invoke some kind of
fast local magnetic diffusion.  In spite of the fact that astrophysical fluids
are turbulent \cite[see][]{arm-95,hor-01,elm-04,mck-07}, the concept of
turbulent magnetic diffusivity \cite[see][]{bla-08},  a popular heuristic
concept used in early  dynamo work, is known to be ill founded
\citep{cat-91, gru-94, vai-92}.  In particular,  it does not address the key question of how
intersecting magnetic fluxes can change their topology.  In an ionized plasma,
Ohmic diffusivity fails by many orders of magnitude to supply the required
magnetic diffusion.

In order for astrophysical dynamos to function smoothly there must be a process
which allows reconnection to proceed at speeds characteristic of local dynamical
velocities.  Since rms fluid velocities are often comparable to the local
Alfv\'en speed, in practice this requirement is indistinguishable from
$V_{rec}\sim V_A$.  This is called fast reconnection, meaning that it does not
depend on resistivity or depends on the resistivity logarithmically
\cite[see][]{par-79}.  There is also direct evidence for fast reconnection from studies of
solar flares \citep{yok-95,inn-97,pag-08}.

The idea that there could be some way of producing fast magnetic reconnection
even in highly conducting fluids is not new \citep{mof-78,kra-80}, but early
models of reconnection \citep[see][e.g.]{par-57, swe-58} using realistic
astrophysical temperatures and densities gave a very slow reconnection rate, $\sim
V_A S^{-1/2}$, where $S\equiv \eta/(V_A L)$ is the Lundquist number, $\eta$ is
the resistivity, and $L$ is the size of the current sheet.  It was \cite{pet-64}
who for the first time introduced a model for fast reconnection with a rate
proportional to $(\log S)^{-1}$.  Subsequent numerical simulations and
theoretical analyses  have shown that the Petschek reconnection rate is only
attainable in very restricted circumstances.  For instance, a modified version
can stably persist in a collisionless plasma  \citep[see][e.g.]{dra-06a}.  This
means that the length of the current sheet should not exceed approximately 50
electron mean free paths \citep{uzd-06,yam-06}.  This condition cannot be
satisfied in many astrophysical environments, e.g. in the interstellar medium
\citep{vis-99}. In a collisional plasma the X-point region required for Petschek
reconnection will collapse to the Sweet-Parker geometry for large $S$
\citep{bis-96}.

The failure of the Petschek model has increased interest in the role of
turbulence in reconnection.  This interest has been further stimulated by the
fact that the turbulence is ubiquitous in astrophysical environments where
reconnection occurs, e.g. the ISM, stars, the Sun and accretion disks
\citep{ruz-88}.  The idea that turbulence can affect reconnection has a long
history, although usually studied in two-dimensions (2D) \citep{pri-00}. Several
researchers have approached this problem numerically, e.g. 
\cite{mat-85,mat-86,fan-04,fan-05,ser-09,lou-09}.  They found that it was
possible to get many features expected from reconnection theory, i.e. large and
small-scale magnetic islands, fluid jetting, current filamentation and that the
maximum reconnection speed was higher for more powerful turbulence and exceeded
the Sweet-Parker rate \citep{fan-04, fan-05}.

The most interesting result of their calculations \citep{fan-04, fan-05} was
that under the influence of turbulence they observed a two-step process of
magnetic reconnection: beginning with a slow Sweet-Parker mechanism and changing
later to a faster reconnection state that they identified with the Petscheck process.
In a similar manner, the most recent paper \citep{lou-09} also observed that the
presence of turbulence significantly enhances the reconnection rate.  They also
found that for a given value of diffusion and above a critical value of the
turbulent injection power the reconnection process accelerates substantially.
However, in all these simulations the authors used periodic boundary conditions,
which prevents inflow or outflow and conserves the total magnetic flux.  This
made the actual reconnection rates difficult to evaluate.  In particular,  it was
impossible to calculate the average reconnection rate, which is important in view
of the large  fluctuations induced by turbulence.  In addition, one can
argue that the reconnection rate in these simulations is influenced by the
conservation of the total magnetic helicity \citep{bla-00}.  These difficulties
can be avoided by using boundary conditions that allow the inflow and outflow of
plasma and magnetic flux \citep{kow-09}.

More recently interest in the magnetic reconnection process has moved from
2D magnetic configurations to more realistic and generic
three-dimensional ones.  In particular, \cite{laz-99} (LV99) and \cite{laz-00}
proposed that in three dimensions a stochastic magnetic field component can
dramatically enhance reconnection rates, leading to reconnection speeds
comparable to the local turbulent velocity.  Their model is based on the
Sweet-Parker reconnection scheme, with a long narrow current sheet between two
regions of dramatically different polarizations but similarly strong magnetic
fields, but includes the effects of turbulence and substructure in the magnetic
field.  This has two principal effects.  First,  in three dimensions many
independent patches of magnetic field come into contact with the current sheet
and undergo reconnection.  Second, the outflow of plasma and shared magnetic
flux happens not over a microscopically narrow region determined by Ohmic
diffusion, but through a substantially wider region determined by field
wandering.  Neither effect is present in two dimensions, although the formation
of magnetic islands in two dimensions is roughly similar to the broadening of the outflow.
Together these effects are sufficient to trigger fast magnetic reconnection.  In
this model the reconnection rate does not depend on the Ohmic resistivity, but
is determined only by turbulence, in particular by its strength and injection
scale.  This fast reconnection model has been tested numerically by
\cite{kow-09} using inflow/outflow boundary conditions and using a wide range
of injection scales and power for the turbulence.  These simulations have
confirmed all of the predicted features of the LV99 model, including the
insensitivity to the Lundquist number.

In this paper we return to two dimensional reconnection using the same
inflow/outflow boundary conditions.  We have two objectives in this work. First,
since the explanation for fast reconnection advanced in LV99 was intrinsically
three dimensional, we are interested in examining the effects of dimensionality
on the reconnection rate.  Second, this work constitutes an examination of the
importance of boundary conditions on the two dimensional model and a test of
claims for fast reconnection in two dimensions.  Our numerical model of
turbulent reconnection in the ISM is calculated in a 2D box with open boundary
conditions.  We use a Harris current sheet setup as an initial configuration.
Reconnection develops as a result of an initial vector potential perturbation.
We do not drive reconnection.  We solve 2D non-ideal normalized isothermal MHD
equations numerically while varying the power and scale of turbulence, and the
magnetic resistivity \citep{kow-09}.

The plan of this paper is as follows. In Sect.~\ref{method} we describe
the numerical setup and input parameters,  and we present our 
method to measure the reconnection rate. The results
are discussed in  Sect.~\ref{results}, where we analyse the 
time evolution of models with different power of injecting turbulence.
We present the dependencies of the reconnection rate on: power of turbulence, 
injection scale, viscosity as well  as uniform  and anomalous  resistivities. We also 
check the influence of initial magnetic field configuration, boundary conditions and 
method of driving turbulence. We discuss our results in Sect.~\ref{discussion} 
and give our conclusions in  Sect.~\ref{conclusion}. 
\noindent

\section{Methods}
\label{method}
\subsection{Basic Equations}
\label{sec:basic}

We study the problem of magnetic reconnection in the presence of weak turbulence
using the magnetized fluid approximation governed by the isothermal non-ideal
MHD equations of the form:
\begin{equation}
\frac{\partial \rho}{\partial t} + \nabla \cdot \left( \rho \vc{v} \right) = 0,
\label{eq:contin}
\end{equation}
\begin{equation}
\frac{\partial \rho \vc{v}}{\partial t} + \nabla \cdot \left( \rho \vc{v} \vc{v} + P_{\ast} \vc{I} - \vc{B} \vc{B} \right) = \vc{f} + \nu \Delta \vc{v},
\label{eq:motion}
\end{equation}
\begin{equation}
\frac{\partial \vc{A}}{\partial t} = \vc{v} \times \vc{B} - \eta \nabla \times \vc{B},
\label{eq:induction}
\end{equation}
where $\rho$ is the gas density, $\vc{v}$ is the fluid velocity, $\vc{A}$ is the
magnetic vector potential, $\vc{B} \equiv \nabla \times \vc{A}$ is the magnetic
field, $P_{\ast} = c_s^2 \rho + B^2 / 8 \pi$ is the total pressure, $\vc{I}$ is
the Kronecker delta, $c_s$ is the isothermal speed of sound, $\eta$ is the
resistivity coefficient, $\nu$ is the viscosity, $\vc{f} = \rho \vc{a}$
represents the forcing term, and $\vc{a}$ is a random acceleration.

We solve the MHD equations using the same code as in \cite{kow-09} based on the
following methods: a higher-order shock-capturing Godunov-type scheme, the
essentially non oscillatory (ENO) spatial reconstruction
\cite[see][]{lon-00,del-03}, a multi-state Harten-Lax-van Leer (HLLD)
approximate Riemann solver for isothermal MHD equations \citep{mig-07} and
Runge-Kutta (RK) time integration \citep[see][e.g.]{del-03}.  The choice of HLLD
Riemann solver guarantees a good solution for the Alfv\'en wave propagation,
which is important in our model, since most of the kinetic energy is transported
through Alfv\'en waves \cite[see][]{kow-09}.  The divergence of the magnetic
field must be kept zero everywhere at all times ($\nabla \cdot \vc{B}=0$).  To
satisfy this condition the field interpolated constraint transport (CT) scheme
based on the staggered grid is used \cite[see][e.g.]{lon-00}.

Some selected simulations that we perform include anomalous resistivity modeled
as
\begin{equation}
\eta = \eta_u + \eta_a \left( \frac{| \vc{j} |}{j_{crit}} - 1 \right) H \left( \frac{| \vc{j} |}{j_{crit}} \right),
\end{equation}
where $\eta_u$ and $\eta_a$ describe uniform and anomalous resistivity
coefficients, respectively, $j_{crit}$ is the critical level of the absolute
value of current density above which the anomalous effects start to work, and
$H$ is a step function.  However, for most of our simulations $\eta_a = 0$.

\subsection{Initial and Boundary Conditions}
\label{sec:initial}

\begin{figure}
\center
\includegraphics[width=0.45\textwidth]{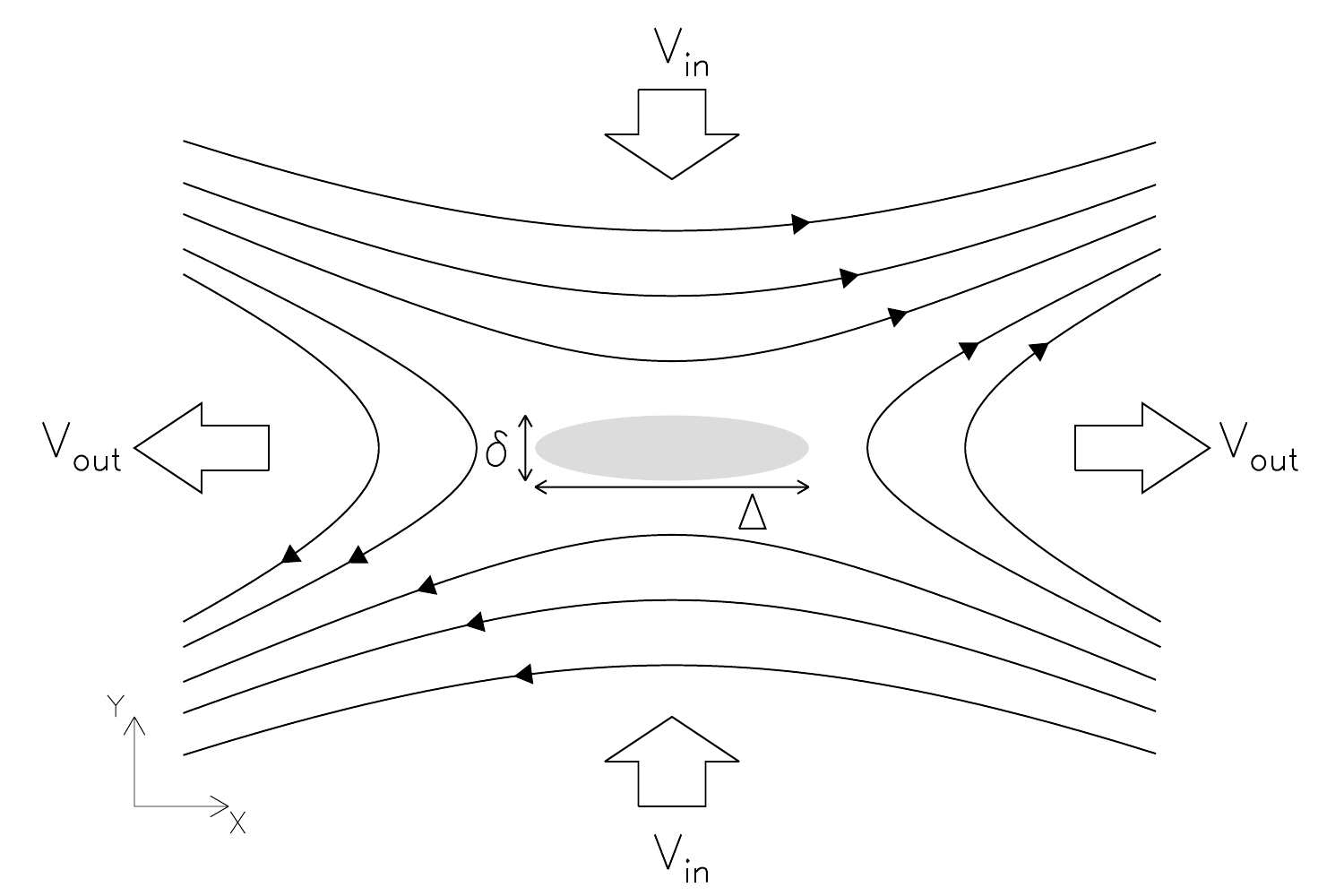}
\caption{2D magnetic field configuration in our problem.  The grey area
describes the diffusion region where the incoming field lines reconnect.  The
longitudinal and transverse scales of the diffusion region are described by the
parameters $\Delta$ and $\delta$, respectively.  We use inflow and outflow
boundary conditions at X and Y directions, respectively. \label{fig:setup}}
\end{figure}

We numerically investigate the turbulent magnetic reconnection model in 2D in
a computational box of size $L_x \times L_y$, where $L_x=1$ and $L_y=2$ with
a spatial resolution of 1024$\times$ 2048 grid zones in the $x$ and $y$
directions, respectively. Figure~\ref{fig:setup} shows a 2D visualization of the
reconnection problem setup.  The domain contains two regions of oppositely
directed magnetic lines separated by a diffusion region with a thickness
$\delta$ and a length $\Delta$, where the magnetic lines reconnect and the product
of this process is ejected along the X direction with a speed $V_{out}$ (see
Figure~\ref{fig:setup}).  The initial magnetic field configuration is described by
a Harris current sheet $B_x(x,y) = B_{x0} \tanh(y / \theta)$, where $B_{x0}$ is
the initial strength of the anti-parallel magnetic field component.  The
Sweet-Parker reconnection is triggered by a small initial perturbation of the
vector potential $\delta \vc{A}(x,y) = \delta B_{0} \cos(2\pi x)
\exp[-(y/d)^2]$, where $d$ and $\delta B_{0}$ denote the thickness of the
perturbed region and the strength of the initial perturbation, respectively. The
initial setup is completed by setting the density profile from the condition of
the uniform total pressure $P_{\ast}(t=0,x,y) = \mathrm{const}$ and setting the
initial velocity to zero everywhere.

The code \citep{kow-09} uses  dimensionless units.  The velocity and
magnetic field are expressed in units of the characteristic Alfv\'{e}n speed
$V_A \equiv |B_{x0}|/\sqrt{\rho_0}$, where the initial anti-parallel component of
magnetic field $B_{x0}=1$ and the density $\rho_0=1$ far from the diffusion region
so that $V_A = 1$ far from the diffusion region  in all models.  
Time is measured in units of the Alfv\'{e}n transit
time $t_A \equiv 1 / V_A$.  The speed of sound is set to $c_s=4$ (the plasma
$\beta \equiv p / p_{mag}$ is 32.0 for all models).  We vary the resistivity
coefficient $\eta_u$ between values $3\cdot10^{-4}$ and $5\cdot10^{-3}$
(in dimensionless units).  In the models which  include anomalous effects, we
vary the anomalous resistivity coefficient $\eta_a$ between 0.0 and
$3\cdot10^{-3}$.  The parameters describing the initial perturbation are set to
$\delta B_{0} = 0.05$ and $d = 0.1$.

We use outflow boundary conditions along the X direction and inflow boundary
conditions along the Y direction setting the normal derivatives of the fluid
variables (density and momentum) to zero.  In the treatment of vector potential
$\vc{A}$ at the boundary we set its components transverse to the considered
boundary using the first order extrapolation, while the normal derivative of the
normal component is set to zero.  This guarantees that all waves generated in
the system are free to leave the box without significant reflections.  For a more
detailed description of these boundary conditions, including their advantages
and drawbacks, we refer the reader to \cite{kow-09}.

\subsection{Model of Turbulence}
\label{sec:forcing}

In order to drive turbulence in our model we follow \cite{kow-09} and use the  method
proposed by \cite{alv-99}. The forcing is driven across  a specified distribution of wave vectors.
Here we use a Gaussian profile around a shell in Fourier space with a radius
which determines the injection scale $l_{inj}$. The forcing is
random in time, and therefore is uncorrelated  with any of the time scales
of the turbulent flow. For the same reason the power input  is defined
purely by the force-velocity correlation. The driving is completely
solenoidal and does not directly produce density fluctuations.

Turbulence is introduced at a given injection scale and grows gradually in time
until it reaches the desired amplitude corresponding to the turbulent power
$P_{inj}$. We drive  turbulence to its saturation level over one Alfv\'{e}nic time from
$t_{beg}=9$ to $t_{end}=10$. According to the LV99 model, the injection scale
and turbulent power determine the rate of reconnection in 3D. Thus, in our model
we test this correlation in 2D by changing these properties of turbulence.

\subsection{Input Parameters}
\begin{table*}
\caption{List of models}
\label{tab:models}
\centering          
\begin{tabular}{c|cccccccccc}      
\hline\hline                      
Name&$\eta_u$ [$10^{-3}$]&$\eta_a$ [$10^{-3}$&$\nu$ [$10^{-3}$] 
&$P_{inj}$&$k_{inj}$&$\Delta k_{inj}$&$N_f$&$\tilde{v}_f$   
&$\langle | \vc{v} | \rangle$\\ 
\hline                    
PD & 0.5 & 0.0 & 0.0 & 0.005 & 12 & 0.5 &  68 & 0.00006 & 0.03983 \\
   & 0.5 & 0.0 & 0.0 & 0.01  & 12 & 0.5 &  68 & 0.00009 & 0.04912 \\
   & 0.5 & 0.0 & 0.0 & 0.05  & 12 & 0.5 &  68 & 0.00019 & 0.08401 \\
   & 0.5 & 0.0 & 0.0 & 0.1   & 12 & 0.5 &  68 & 0.00028 & 0.10645 \\
   & 0.5 & 0.0 & 0.0 & 0.5   & 12 & 0.5 &  68 & 0.00062 & 0.21685 \\
   & 0.5 & 0.0 & 0.0 & 1.0   & 12 & 0.5 &  68 & 0.00088 & 0.28657 \\
\hline
SD & 0.5 & 0.0 & 0.0 & 0.1   &  8 & 0.5 &  48 & 0.00033 & 0.11618 \\
   & 0.5 & 0.0 & 0.0 & 0.1   & 12 & 0.5 &  68 & 0.00028& 0.10645 \\
   & 0.5 & 0.0 & 0.0 & 0.1   & 16 & 0.5 & 112 & 0.00029 & 0.08927 \\
   & 0.5 & 0.0 & 0.0 & 0.1   & 20 & 0.5 & 112 & 0.00023 & 0.06626 \\
   & 0.5 & 0.0 & 0.0 & 0.1   & 24 & 1.0 &  36 & 0.00042 & 0.05534 \\
   & 0.5 & 0.0 & 0.0 & 0.1   & 28 & 1.0 &  32 & 0.00058 & 0.04633 \\
   & 0.5 & 0.0 & 0.0 & 0.1   & 32 & 1.0 &  32 & 0.00056 & 0.03997 \\
\hline
RD & 0.3 & 0.0 & 0.0 & 0.01  & 12 & 1.0 &  68 & 0.00009 & 0.05734 \\
   & 0.5 & 0.0 & 0.0 & 0.01  & 12 & 1.0 &  68 &  0.00009 & 0.04912 \\
   & 0.6 & 0.0 & 0.0 & 0.01  & 12 & 1.0 &  68 &  0.00009 & 0.04471 \\
   & 0.7 & 0.0 & 0.0 & 0.01  & 12 & 1.0 &  68 & 0.00009 & 0.04889 \\
   & 0.8 & 0.0 & 0.0 & 0.01  & 12 & 1.0 &  68 & 0.00009 & 0.05405 \\
   & 0.9 & 0.0 & 0.0 & 0.01  & 12 & 1.0 &  68 & 0.00009 & 0.04813 \\
   & 1.0 & 0.0 & 0.0 & 0.01  & 12 & 1.0 &  68 & 0.00009 & 0.05481 \\
   & 2.0 & 0.0 & 0.0 & 0.01  & 12 & 1.0 &  68 & 0.00009 & 0.05542 \\
   & 3.0 & 0.0 & 0.0 & 0.01  & 12 & 1.0 &  68 & 0.00009 & 0.06214 \\
   & 4.0 & 0.0 & 0.0 & 0.01  & 12 & 1.0 &  68 & 0.00009 & 0.06774 \\
   & 5.0 & 0.0 & 0.0 & 0.01  & 12 & 1.0 &  68 & 0.00009 & 0.07407 \\
   & 0.3 & 0.0 & 0.0 & 0.1   & 12 & 1.0 &  68 & 0.00028 & 0.11583 \\
   & 0.5 & 0.0 & 0.0 & 0.1   & 12 & 1.0 &  68 & 0.00028 & 0.10649 \\
   & 0.6 & 0.0 & 0.0 & 0.1   & 12 & 1.0 &  68 & 0.00028 & 0.10763 \\
   & 0.7 & 0.0 & 0.0 & 0.1   & 12 & 1.0 &  68 & 0.00028 & 0.10122 \\
   & 0.8 & 0.0 & 0.0 & 0.1   & 12 & 1.0 &  68 & 0.00028 & 0.10221 \\
   & 0.9 & 0.0 & 0.0 & 0.1   & 12 & 1.0 &  68 & 0.00028 & 0.10036 \\
   & 1.0 & 0.0 & 0.0 & 0.1   & 12 & 1.0 &  68 & 0.00028 & 0.09854 \\
   & 2.0 & 0.0 & 0.0 & 0.1   & 12 & 1.0 &  68 & 0.00028 & 0.08844 \\
   & 3.0 & 0.0 & 0.0 & 0.1   & 12 & 1.0 &  68 & 0.00028 & 0.08075 \\
   & 4.0 & 0.0 & 0.0 & 0.1   & 12 & 1.0 &  68 & 0.00028 & 0.07236 \\
   & 5.0 & 0.0 & 0.0 & 0.1   & 12 & 1.0 &  68 & 0.00028 & 0.06987 \\
\hline
AD & 0.5 & 0.0 & 0.0 & 0.1   & 12 & 1.0 &  68 & 0.00028 & 0.10645 \\
   & 0.5 & 0.5 & 0.0 & 0.1   & 12 & 1.0 &  68 & 0.00028 & 0.10534 \\
   & 0.5 & 1.0 & 0.0 & 0.1   & 12 & 1.0 &  68 & 0.00028 & 0.11071 \\
   & 0.5 & 2.0 & 0.0 & 0.1   & 12 & 1.0 &  68 & 0.00028 & 0.11252 \\
   & 0.5 & 3.0 & 0.0 & 0.1   & 12 & 1.0 &  68 & 0.00028 & 0.11465 \\
\hline
VD & 0.5 & 0.0 & 0.08 & 0.1  & 12 & 1.0 &  68 & 0.00028 & 0.10372 \\
   & 0.5 & 0.0 & 0.09 & 0.1  & 12 & 1.0 &  68 & 0.00028 & 0.09623 \\
   & 0.5 & 0.0 & 0.10 & 0.1  & 12 & 1.0 &  68 & 0.00028 & 0.09041 \\
   & 0.5 & 0.0 & 0.20 & 0.1  & 12 & 1.0 &  68 & 0.00028 & 0.08472 \\
   & 0.5 & 0.0 & 0.30 & 0.1  & 12 & 1.0 &  68 & 0.00028 & 0.08366 \\
   & 0.5 & 0.0 & 0.50 & 0.1  & 12 & 1.0 &  68 & 0.00028 & 0.07317 \\
   & 0.5 & 0.0 & 0.60 & 0.1  & 12 & 1.0 &  68 & 0.00028 & 0.06957 \\
   & 0.5 & 0.0 & 0.80 & 0.1  & 12 & 1.0 &  68 & 0.00028 & 0.06868 \\
   & 0.5 & 0.0 & 1.00 & 0.1  & 12 & 1.0 &  68 & 0.00028 & 0.06019 \\
   & 0.5 & 0.0 & 2.00 & 0.1  & 12 & 1.0 &  68 & 0.00028 & 0.04371 \\
   & 0.5 & 0.0 & 3.00 & 0.1  & 12 & 1.0 &  68 & 0.00028 & 0.03852 \\
   & 0.5 & 0.0 & 4.00 & 0.1  & 12 & 1.0 &  68 & 0.00028 & 0.03394 \\
   & 0.5 & 0.0 & 5.00 & 0.1  & 12 & 1.0 &  68 & 0.00028 & 0.02912 \\
\hline
\end{tabular}
\end{table*}
Our simulations can be divided into five groups. In each of them we analyze the
dependence of the reconnection rate by changing only one of the crucial parameters:
for models PD - the power of turbulence $P_{inj}$, for  models SD - the injection wavenumber $k_f$,
for models VD - viscosity $\nu$, for models RD and AD the uniform $\eta_u$ and anomalous
$\eta_a$ resistivities, respectively. In Table~\ref{tab:models} we list
parameters of all the models presented in this paper.

Among all parameters of the model we list those which vary, i.e. the uniform and
anomalous resistivities, $\eta_u$ and $\eta_a$, respectively, the uniform
viscosity $\nu$, the power of turbulence $P_{inj}$ and its injection wavenumber 
$k_{inj}$ with the half-thickness of the injection shell $\Delta k_{inj}$, the
number of perturbed Fourier components of velocity $N_f$ and the amplitude of
perturbation $\tilde{v}_f$ at the injection scale. In addition, we include the
mean velocity amplitude obtained at the end of each simulation.


\subsection{Reconnection Rate Measure}
In the case of laminar reconnection (Sweet-Parker and Petschek) its rate can be
measured by averaging the inflow velocity $V_{in}$ divided by the Alfv\'en speed
$V_A$ over the inflow boundaries, i.e.
\begin{equation}
\langle V_{in} / V_A \rangle = \frac{1}{2} \int_{x_{min}}^{x_{max}} {dx \left( \left. \frac{v_y}{V_A} \right|_{y=y_{min}} - \left. \frac{v_y}{V_A} \right|_{y=y_{max}} \right)}.
\end{equation}
Since we have two X boundaries, located at $y=y_{min}$ and $y=y_{max}$, we need
to take half of the resulting integral.  This measure works well for laminar
reconnection, when the system is perfectly stable and where the time derivative
of the magnetic flux is zero.  In the presence of turbulence, however, this time
derivative can fluctuate or the turbulence in the center of the box could affect
the flow of the plasma.  In this way we would get a flow of magnetic flux
without the presence of reconnection.  In order to include all effects
contributing to the change of magnetic flux, we use a new more general measure
of the reconnection rate as described in \cite{kow-09}, which in 2D is given by
the following simplified equation:
\begin{eqnarray}
V_{rec} & = & \frac{1}{2|B_{x,\infty}|} \bigg[
 \left. \left( \mathrm{sign}(B_x) E_z \right) \right|_{y_{max}} - \left. \left( \mathrm{sign}(B_x) E_z \right) \right|_{y_{min}} \nonumber \\
 & & - \partial_t \int_{y_{min}}^{y_{max}}{dy |B_x|} \bigg],
\label{rec_new}
\end{eqnarray}
where $|B_{x,\infty}|$ is the asymptotic absolute value of $B_x$ and $E_z$ is
the Z component of the electromotive force.  A more complete discussion of
this new reconnection measure can be found in \cite{kow-09}.

\section{Results}
\label{results}


\subsection{Laminar Reconnection}
\label{laminar}

During the  first stage of our simulations, before we start driving turbulence, the system
evolves  to reach the stationary state of Sweet-Parker reconnection.  After
that the influence of turbulence on the evolution of our system can be studied
precisely.  As mentioned above we start our simulations with a configuration of
the oppositely directed initial magnetic fields with a small magnetic
perturbation.  The perturbation initiates  Sweet-Parker reconnection which
reaches a stationary state in a few Alf\'en time units which lasts until we start
injecting turbulence.

\begin{figure}
\center
\includegraphics[width=\columnwidth]{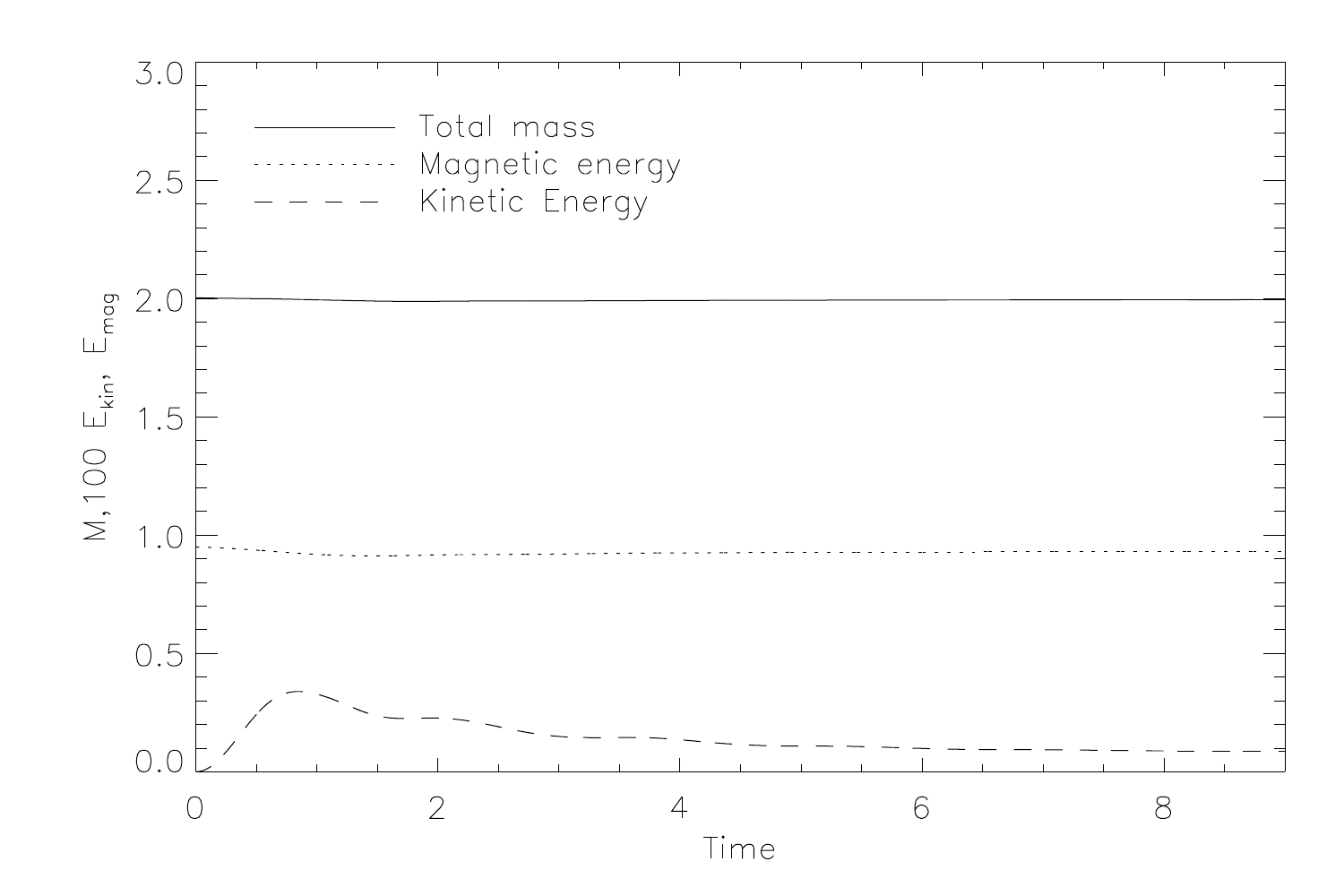}
\caption{Time variations of the total mass $M$ (solid line), magnetic $E_{mag}$
(dashed line) and kinetic $E_{kin}$ (dotted line) energies during the
Sweet-Parker reconnection stage with the uniform resistivity
$\eta_u=5\cdot10^{-4}$.  For clarity  the kinetic
energy $E_{kin}$ has been amplified by a factor of 100. \label{fig:con_sp}}
\end{figure}

\begin{figure}
\center
\includegraphics[width=\columnwidth]{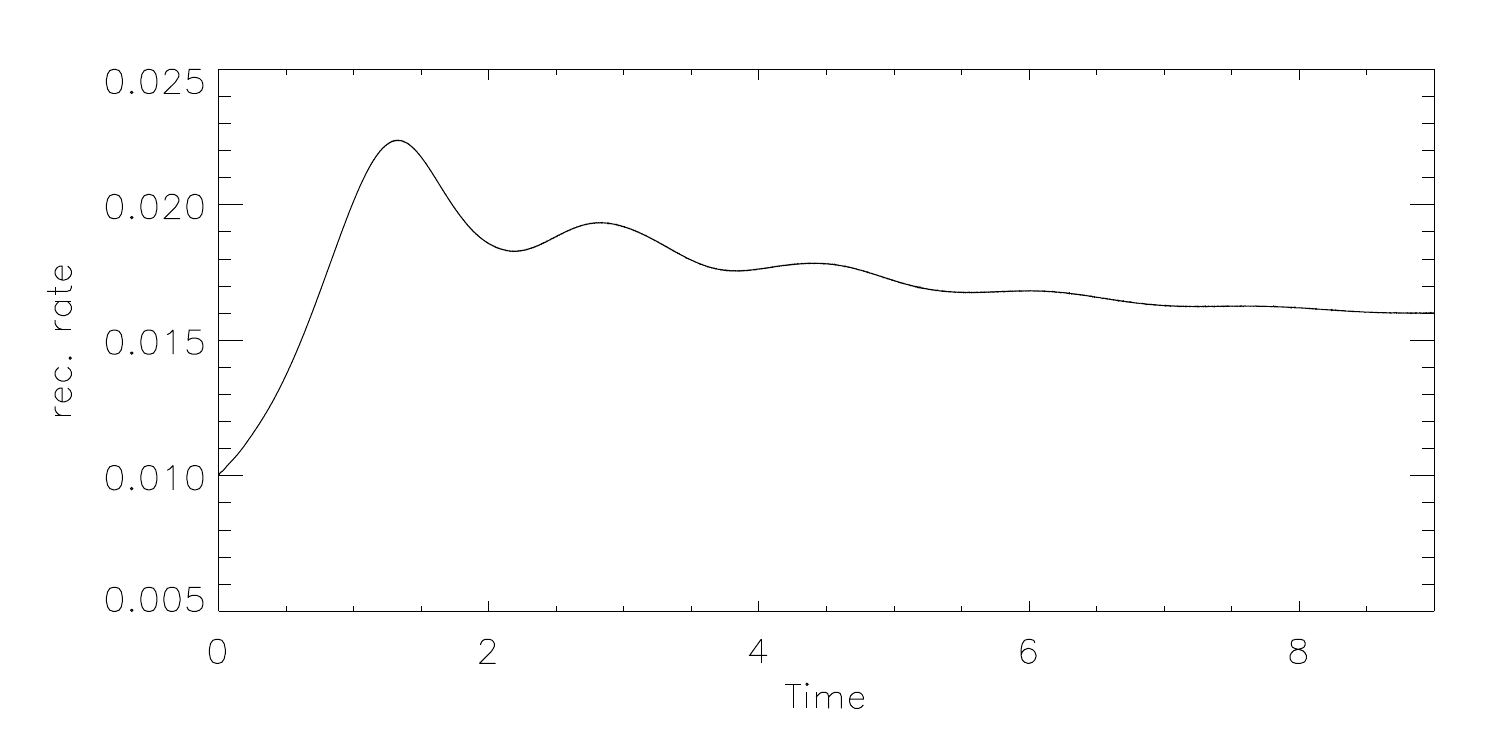}
\caption{Time variation evolution of the reconnection rate during the Sweet-Parker stage
with the uniform resistivity $\eta_u=5\cdot10^{-4}$. \label{fig:vsp}}
\end{figure}

We assume that we get the steady state of Sweet-Parker reconnection when the
total mass, reconnection rate, kinetic and magnetic energies show very small time derivatives.
Figure~\ref{fig:con_sp} shows the evolution of mass,
kinetic and magnetic energy during the laminar reconnection stage.  In the
beginning all these quantities change slightly but then reach almost constant values.
The reconnection rate (Figure~\ref{fig:vsp}) initially increases
until   $t \sim 1.3$, when it starts to oscillate, finally stabilizing  
after a few time units.   
\begin{figure*}
\centering
\includegraphics[width=0.3\textwidth]{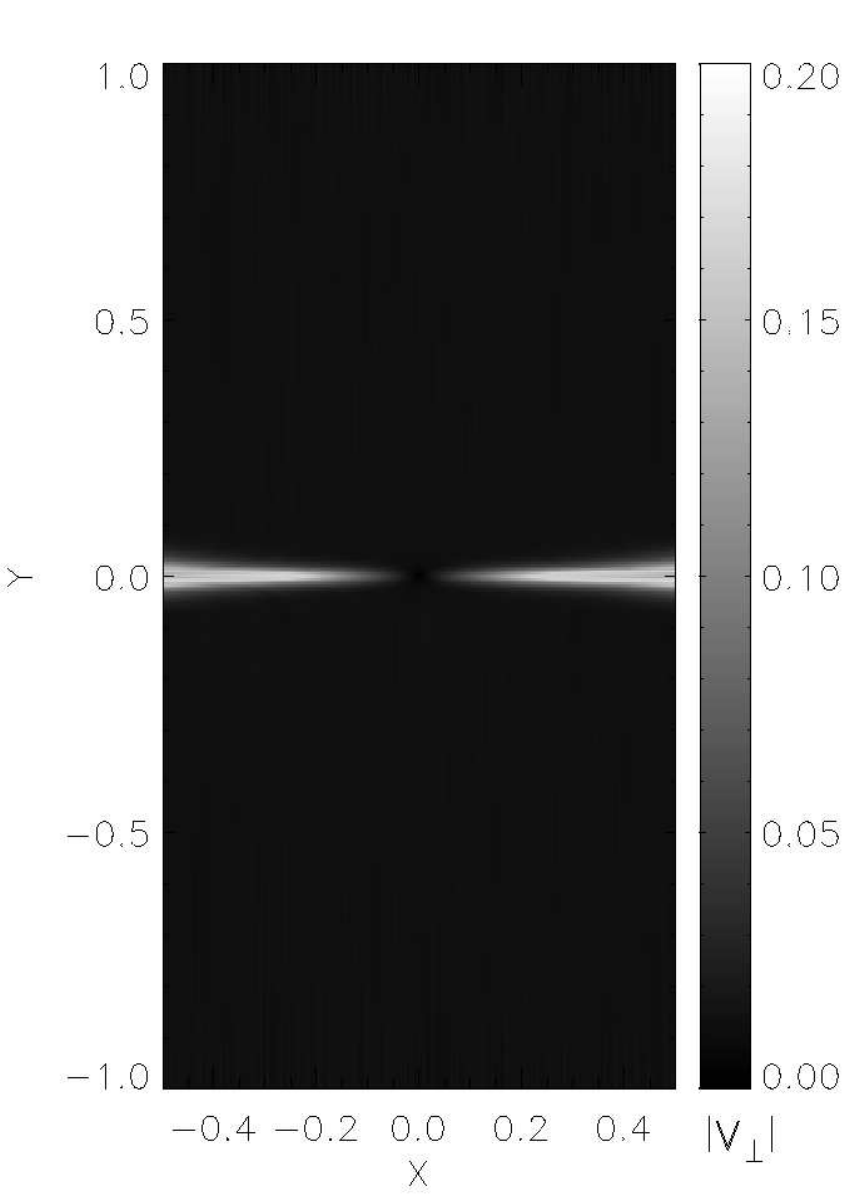}
\includegraphics[width=0.3\textwidth]{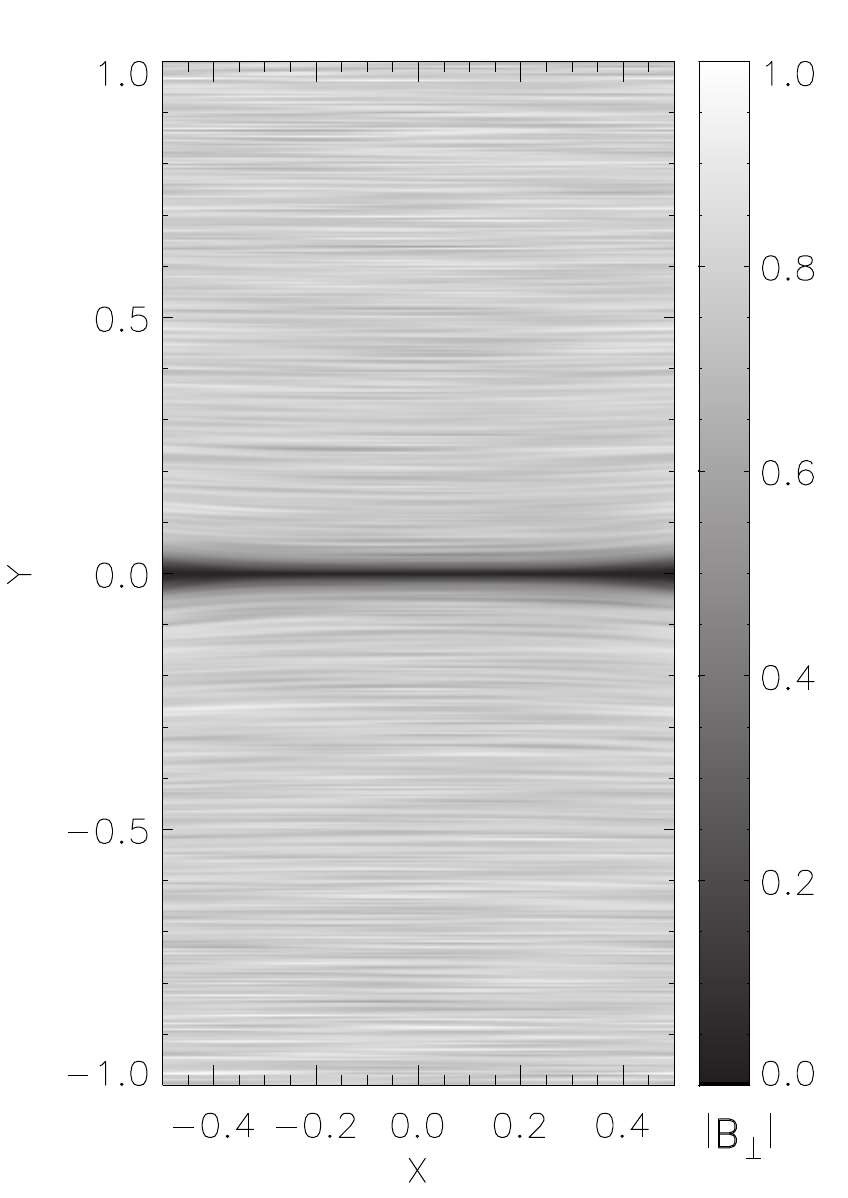}
\includegraphics[width=0.3\textwidth]{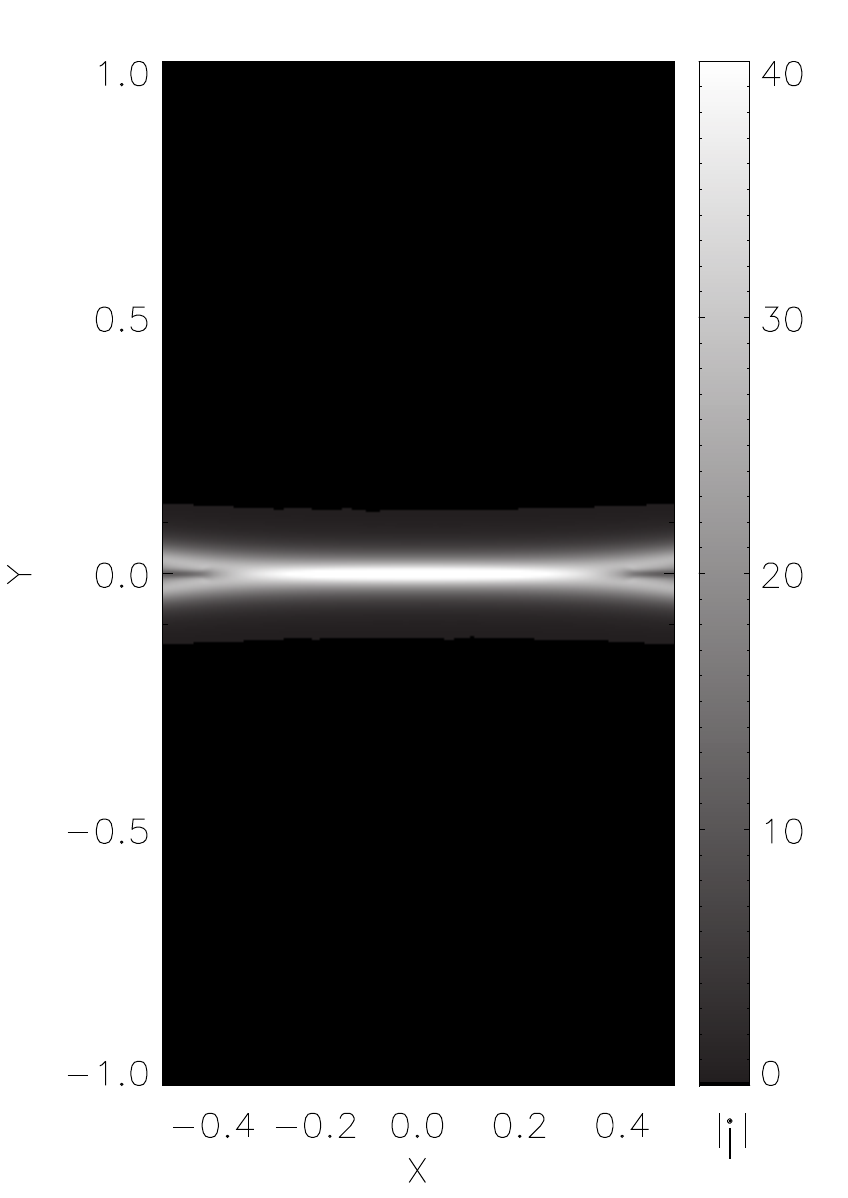}
\caption{Absolute value of the current density distribution (right panel), the
magnetic (middle panel) and velocity field (left panel) topology during the
Sweet-Parker stage for time $t=9$ in a model with the uniform resistivity
$\eta_u=5\cdot10^{-4}$.  Brighter shades correspond to the larger values of the
displayed quantities.  All panels show the moment just before we start injecting
turbulence. \label{fig:sp_top}}
\end{figure*}

Figure~\ref{fig:sp_top} shows the topology of the velocity (left panel) and magnetic
fields (middle panel), and the absolute value of the current density (right panel)
just before we start injecting turbulence. The brighter shades correspond to
large values of the displayed quantities.  The velocity field  and magnetic field
are shown in the form of texture.  The initial oppositely directed magnetic
field lines are transported to the middle of the box. The Y-component of magnetic field emerges from
the current sheet and is ejected near the
midplane through the left and right boundaries.  The reconnection process in the
diffusion region and the ejection of $B_y$ cause the strong outflow of gas
which is clearly visible in the left panel of Figure~\ref{fig:sp_top}.  The system reaches a steady state
despite this loss of mass due to the inflow of gas through the top and bottom of  the computational box.

The absolute value of current density is shown in the right panel of 
Figure~\ref{fig:sp_top}. As expected   the highest values of the
current density appear in the midplane and determine the diffusion region.  The
state of Sweet-Parker reconnection described above is stationary and sufficient
to study the influence of turbulence on the reconnection process.

\subsection{Reconnection in the Presence of Turbulence}
\label{turb_reconn}

In this section we present the influence of turbulence on the reconnection
process.   Turbulence is injected
at time $t=9$ and at a given injection scale $l_{inj}\propto k_f^{-1}$ in the
vicinity of the midplane.  We gradually increase the strength of turbulence during 
one Alfv\'en time, thus at time $t=10$ the power reaches its input value
defined by $P_{inj}$.

\begin{figure}
\center
\includegraphics[width=\columnwidth]{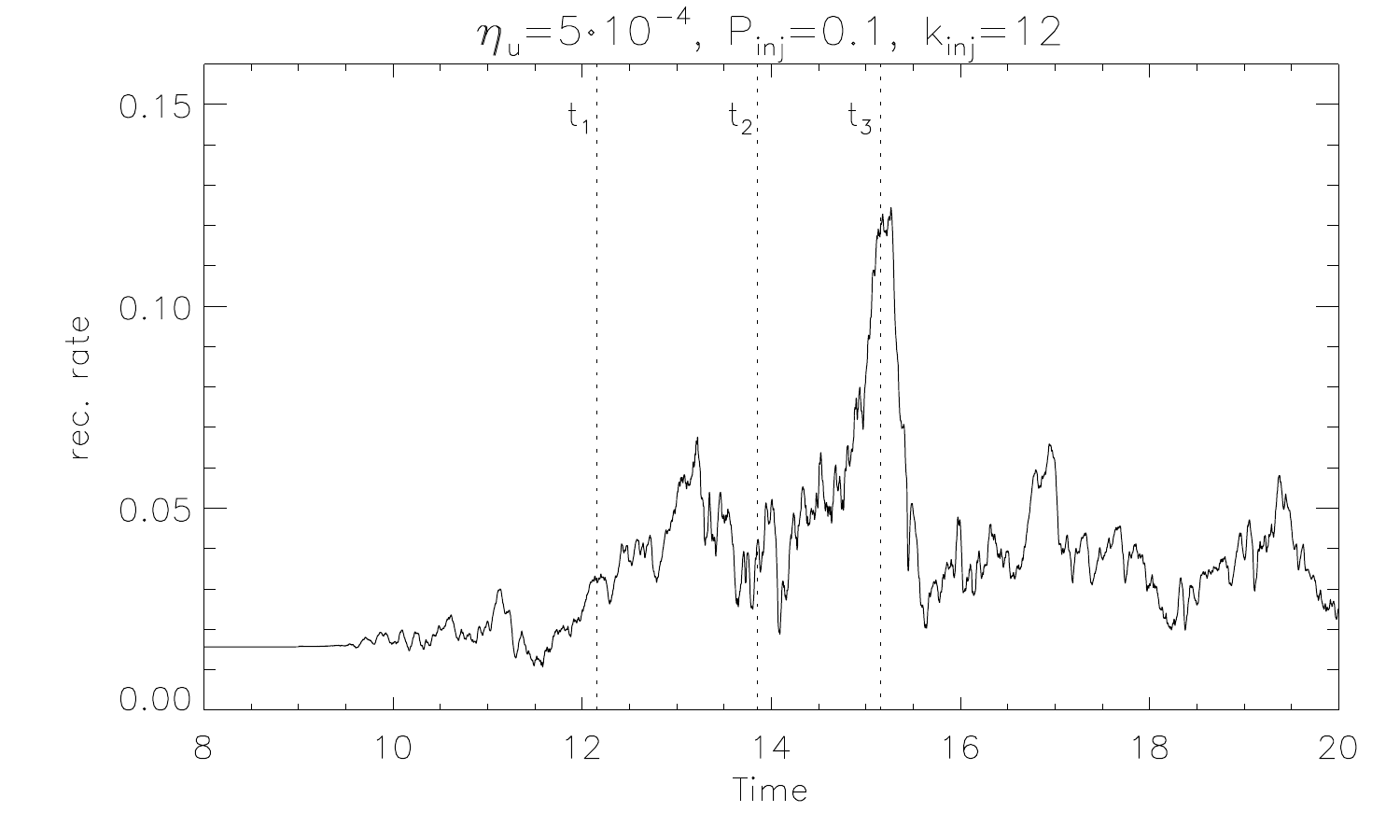}
\caption{The evolution of the reconnection speed for a model with  $P_{inj}=0.1$,  a mean wavenumber of 
the injected turbulence $k_f=12$ and  uniform
resistivity  $\eta_u=5\cdot10^{-4}$.  Marked time steps correspond to those 
presented in Figure~\ref{fig:ev_p0.1}. \label{fig:vr_p0.1}}
\end{figure}

\begin{figure*}
\center
\includegraphics[width=0.3\textwidth]{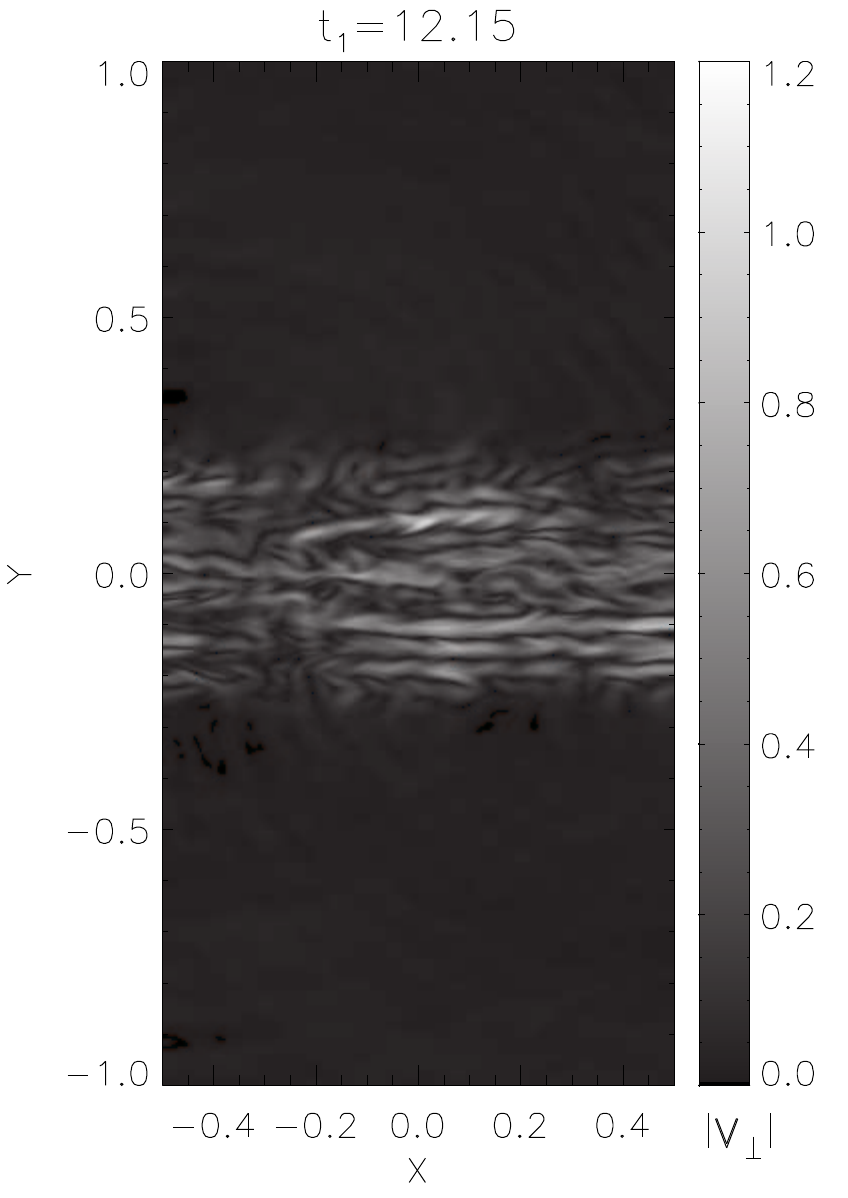}
\includegraphics[width=0.3\textwidth]{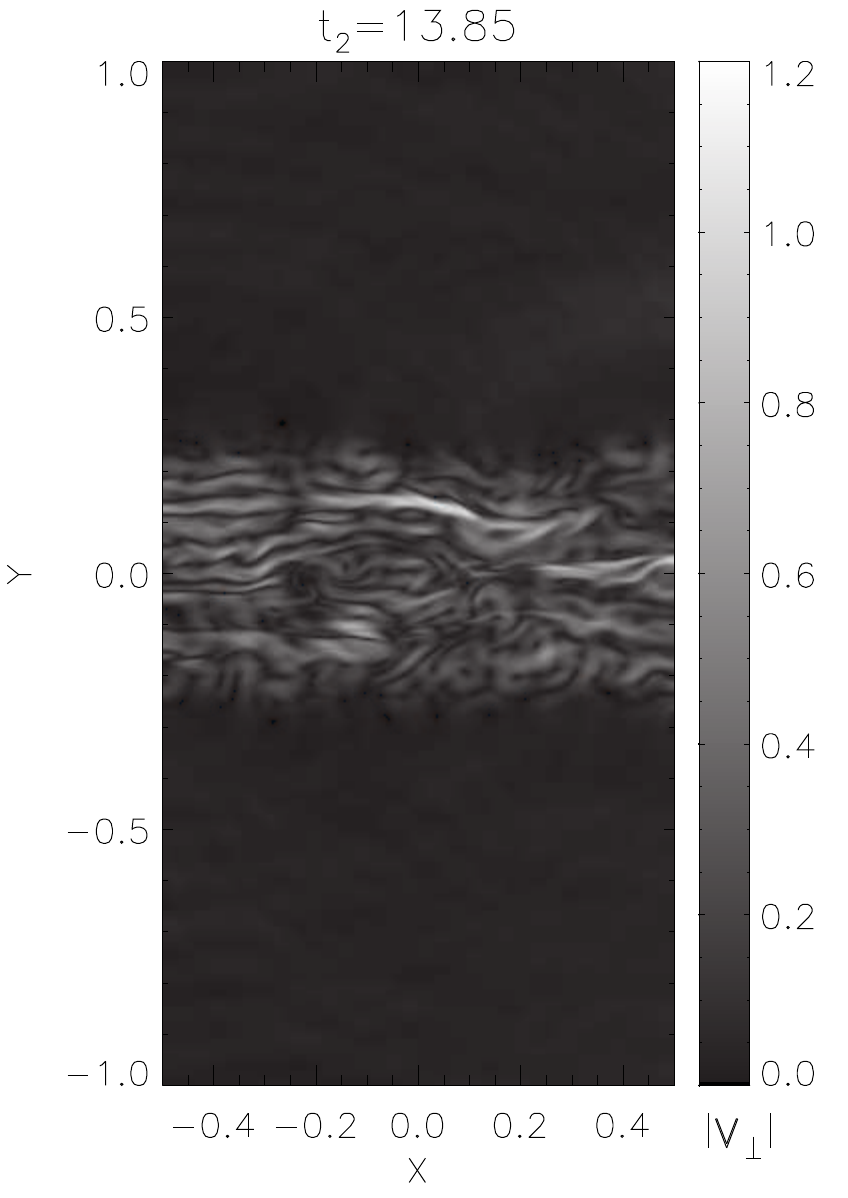}
\includegraphics[width=0.3\textwidth]{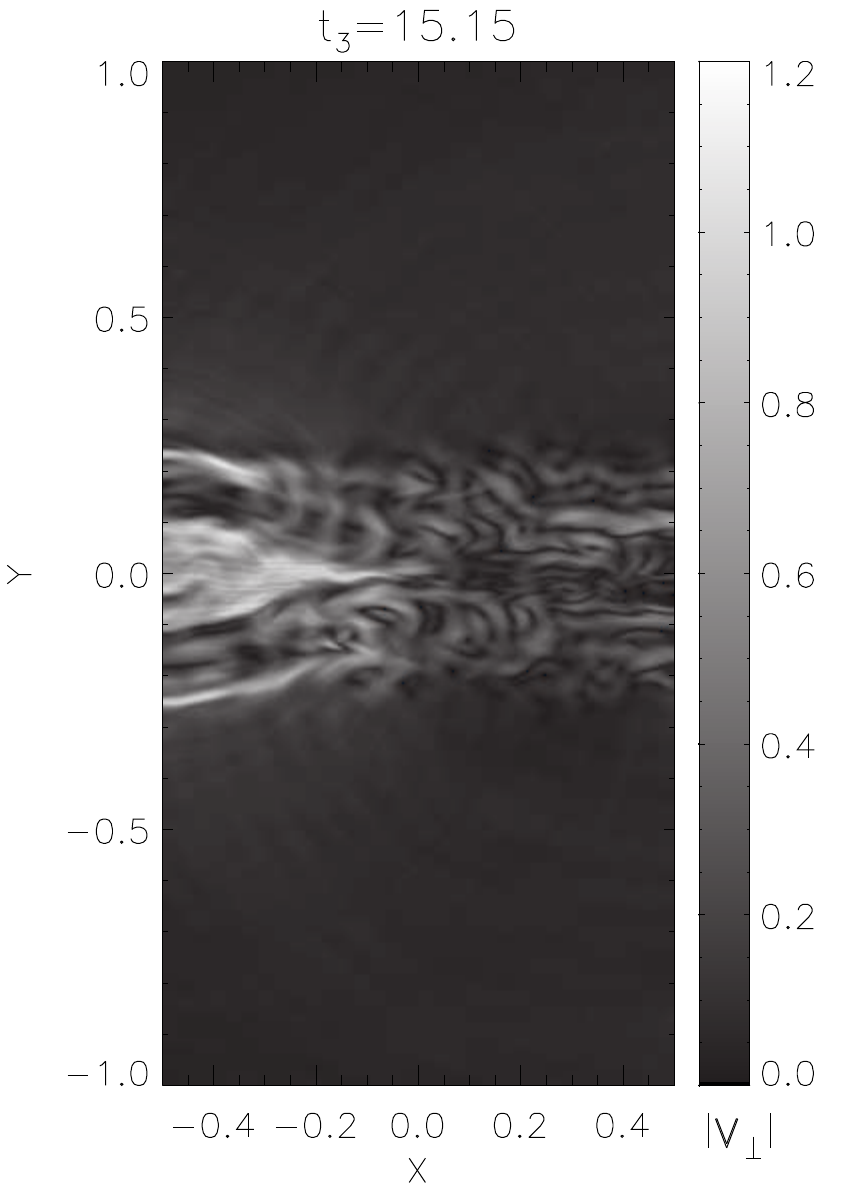} \\
\includegraphics[width=0.3\textwidth]{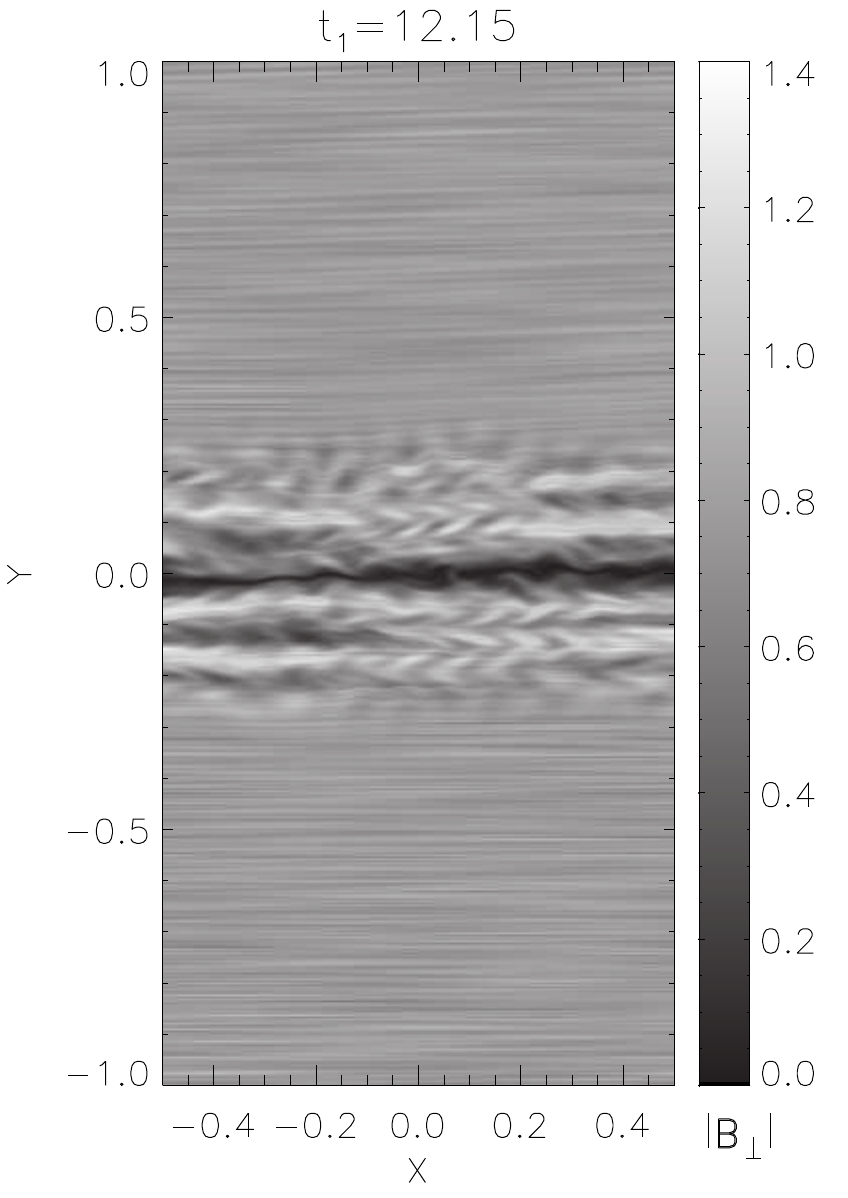}
\includegraphics[width=0.3\textwidth]{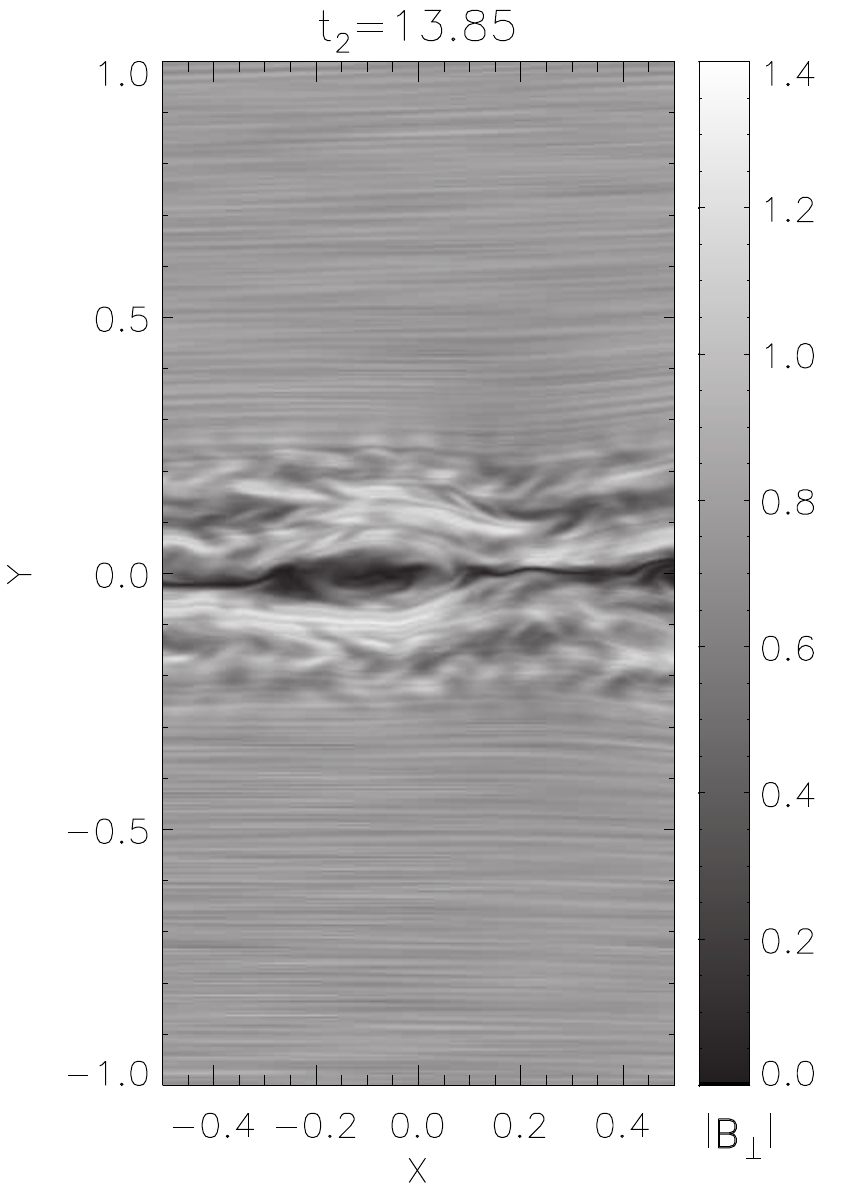}
\includegraphics[width=0.3\textwidth]{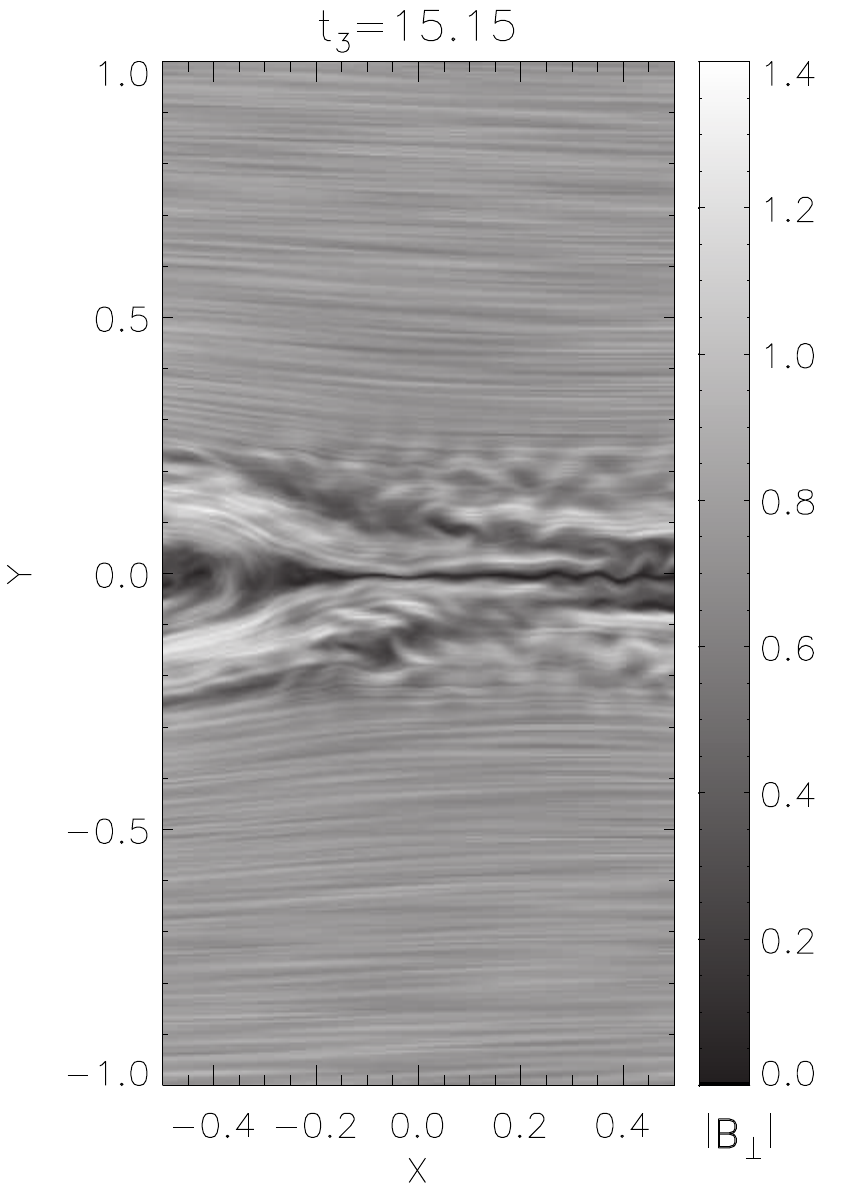} \\
\includegraphics[width=0.3\textwidth]{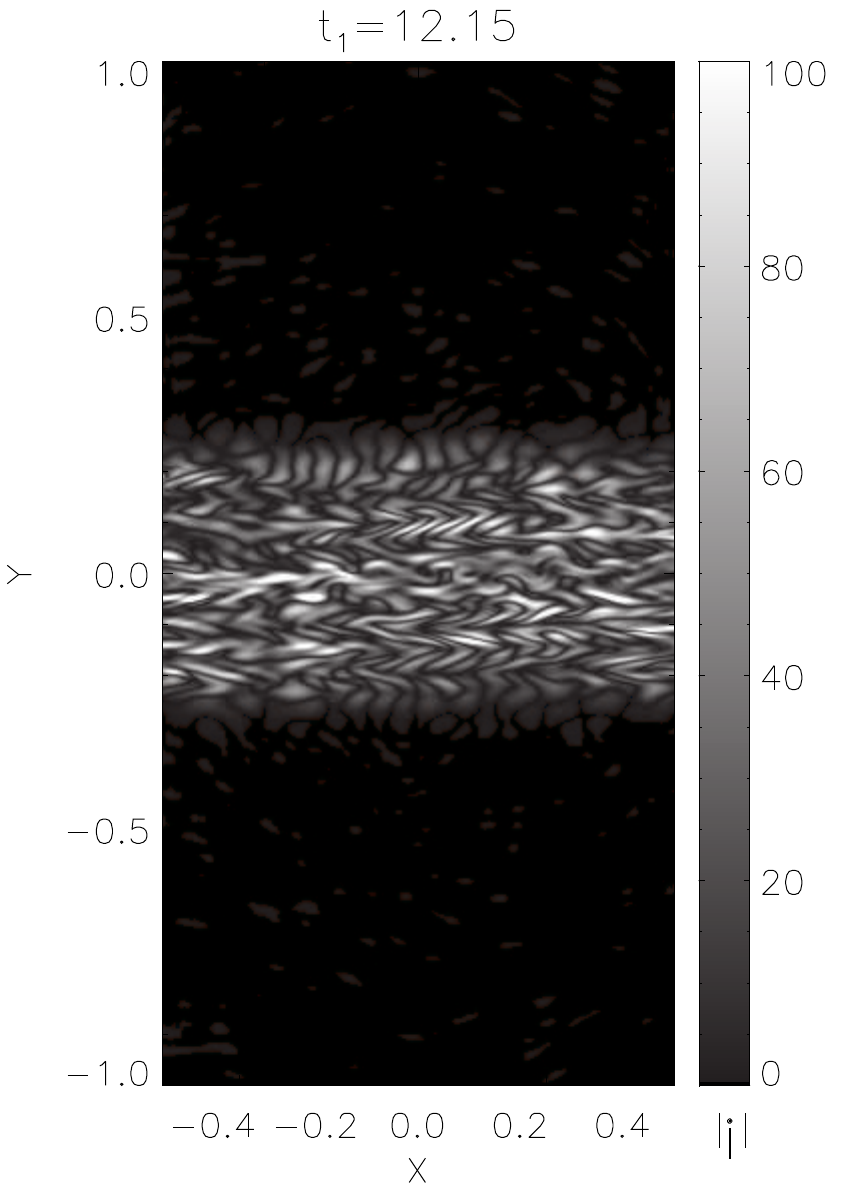}
\includegraphics[width=0.3\textwidth]{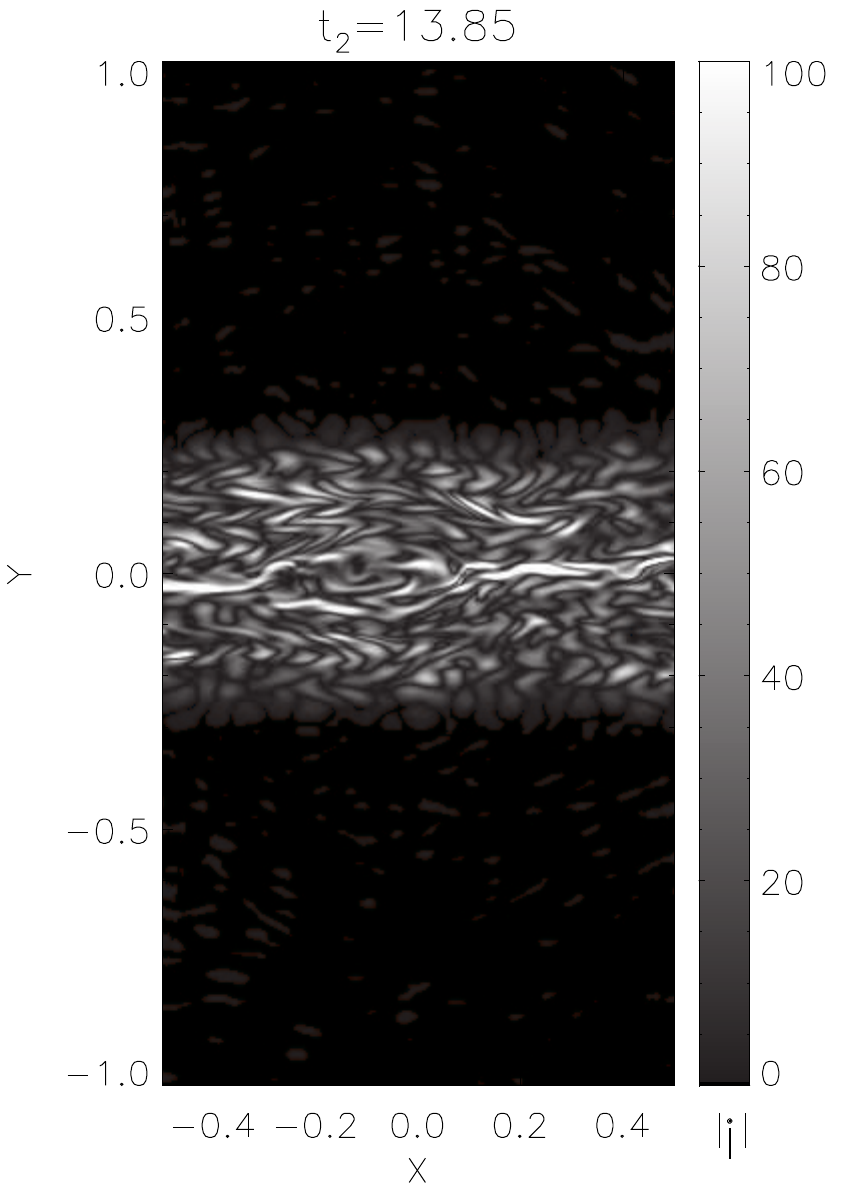}
\includegraphics[width=0.3\textwidth]{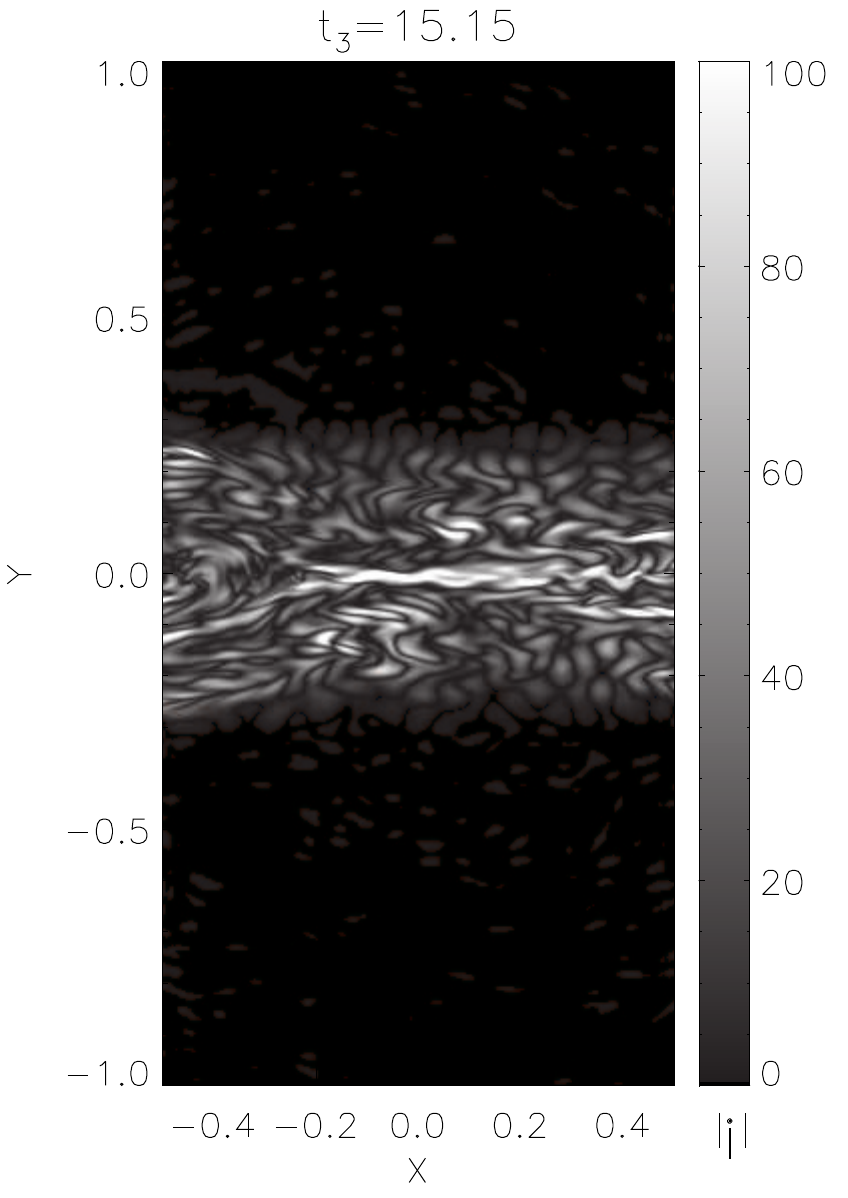}
\caption{Topology of velocity (top panels) and  magnetic field (middle panels)
in the presence of turbulence shown at three times $t_1=12.15$, $t_2=13.85$,
$t_3=15.15$.  In the lower panels we plot the  absolute value of current density at those times. 
The variation of the
reconnection speed $V_r^{TB}$ for this model is presented in Figure~\ref{fig:vr_p0.1}. 
Turbulence is injected with power $P_{inj}=0.5$ and at scale $k=12$.  
The uniform resistivity $\eta_u$ is equal $5\cdot10^{-4}$. The brighter 
shades correspond to the larger values of displayed quantities. \label{fig:ev_p0.1}}
\end{figure*}

Figure~\ref{fig:vr_p0.1} shows the reconnection rate obtained for a model with 
uniform resistivity $\eta_u=5\cdot10^{-4}$, a turbulent power
 $P_{inj}=0.1$ and an injection wavenumber  $k_f=12$. Adding turbulence to
the system results in slight fluctuations of the reconnection rate until
$t=11.5$. Next, the reconnection rate increases significantly until it reaches
the maximum at time $t\sim13$ and then drops. We can  distinguish four such
maxima, which are roughly separated by   two Alfv\'en times  ($t\sim13$,
$t\sim15$, $t\sim17$, $t\sim19$). In~Figure \ref{fig:vr_p0.1} we mark three time
steps corresponding to the same time stages shown in Figure~\ref{fig:ev_p0.1},
where we plot the current density, magnetic and gas velocity fields.

In Figure~\ref{fig:ev_p0.1} (top and middle row) we see that the structure
of the magnetic and velocity fields are considerably different than in the case of
Sweet-Parker reconnection (see Figure~\ref{fig:sp_top}). At   
$t_1=12.15$ (Figure~\ref{fig:ev_p0.1}, first column, top panel) we are injecting
turbulence with the maximum power $P_{inj}=0.1$ in a large volume surrounding the the midplane.
Thus, in this region the velocity field is strongly perturbed and mixed.
Although the topology of the velocity field is very complicated, we can
distinguish the main direction of the velocity fluctuations, which is parallel to
the mean magnetic field.

A smaller number of distinct features in the velocity  field are  pointed
perpendicular to the mean magnetic field. Close to the midplane the  magnetic
field lines change their directions and are substantially reduced in strength. Thus, velocity
fluctuations can bend magnetic field lines in this region.  Another noticeable
difference in comparison to the Sweet-Parker configuration is a significant
change of  the current density distribution. Adding turbulence to the system creates a very
complex configuration of the magnetic field, so that we observe multiple 
reconnection events happening at the same time 
(Figure~\ref{fig:ev_p0.1}, $t_1=12.15$, bottom panel).

The magnetic field configuration looks quite different at the next time step
$t_2=13.85$ (Figure~\ref{fig:ev_p0.1}, second column, middle panel). In the 
vicinity of the midplane we observe the formation of a magnetic island. Also the
velocity field  (Figure~\ref{fig:ev_p0.1}, second column, top panel) is more
mixed in this region. However, the outflow velocity of the gas velocity is quite low, which is
confirmed by the small reconnection rate (Figure~\ref{fig:vr_p0.1}).

As the simulation proceeds a large loop of magnetic field lines moves to the
left boundary (Figure~\ref{fig:ev_p0.1}, $t_3=15.15$, last column). This causes
a powerful outflow of gas, which results in a violent and rapid growth of
the reconnection rate (Figure~\ref{fig:vr_p0.1}).   We see that the maxima of the
velocity field associated with this loop are where the value of
magnetic field is low. The area  of the magnetic island  
$t_2=13.85$ and $t_3=15.15$ is also well defined by the current density
distribution (Figure~\ref{fig:ev_p0.1},  bottom panel, middle and right column).

\begin{figure}
\center
\includegraphics[width=\columnwidth]{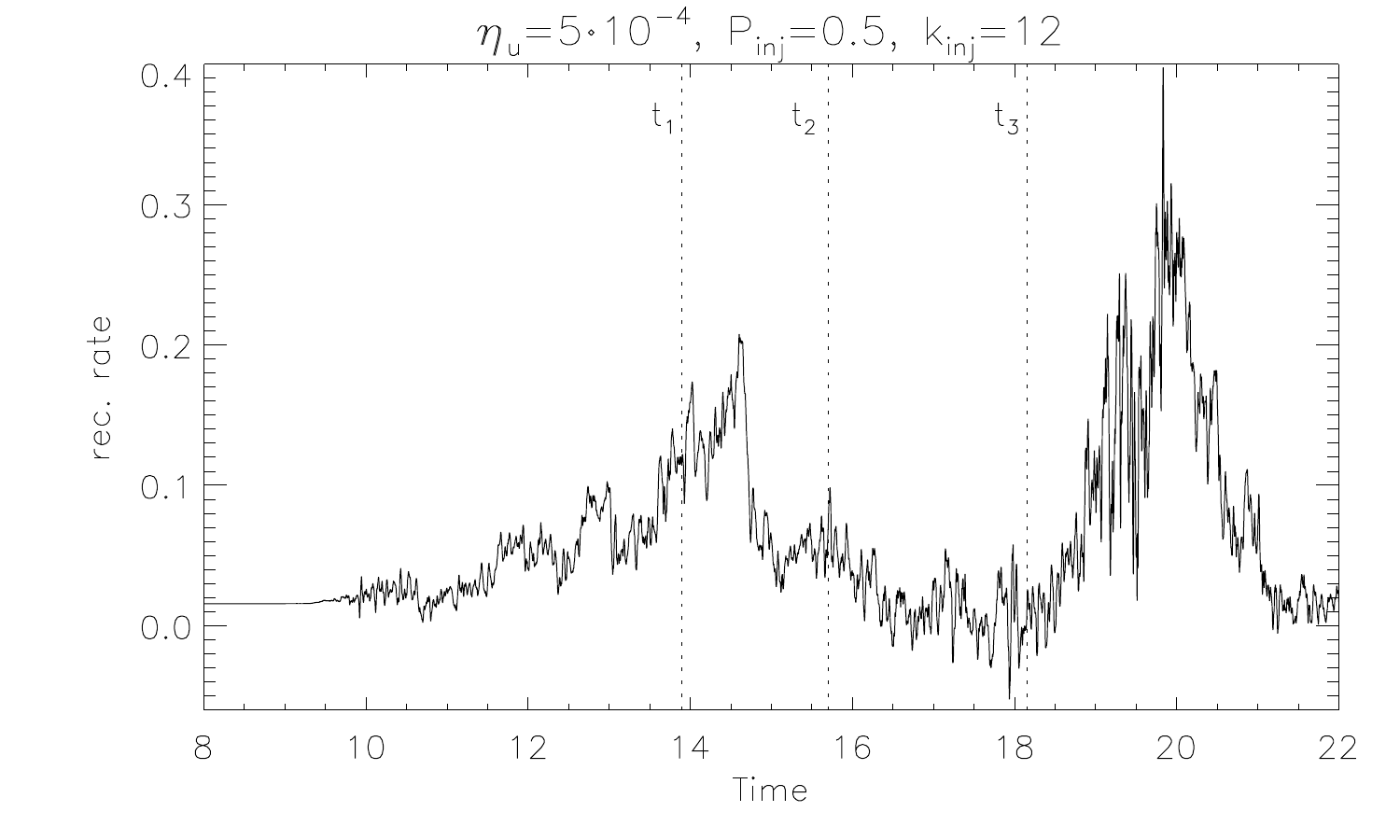}
\caption{The evolution of the reconnection speed for the model with   $P_{inj}=0.5$,   injection scale $k_f=12$ and uniform
resistivity $\eta_u=5\cdot10^{-4}$.  Marked times correspond to those shown
in Figure~\ref{fig:ev_p0.5}. \label{fig:vr_p0.5}}
\end{figure}

\begin{figure*}
\center
\includegraphics[width=0.3\textwidth]{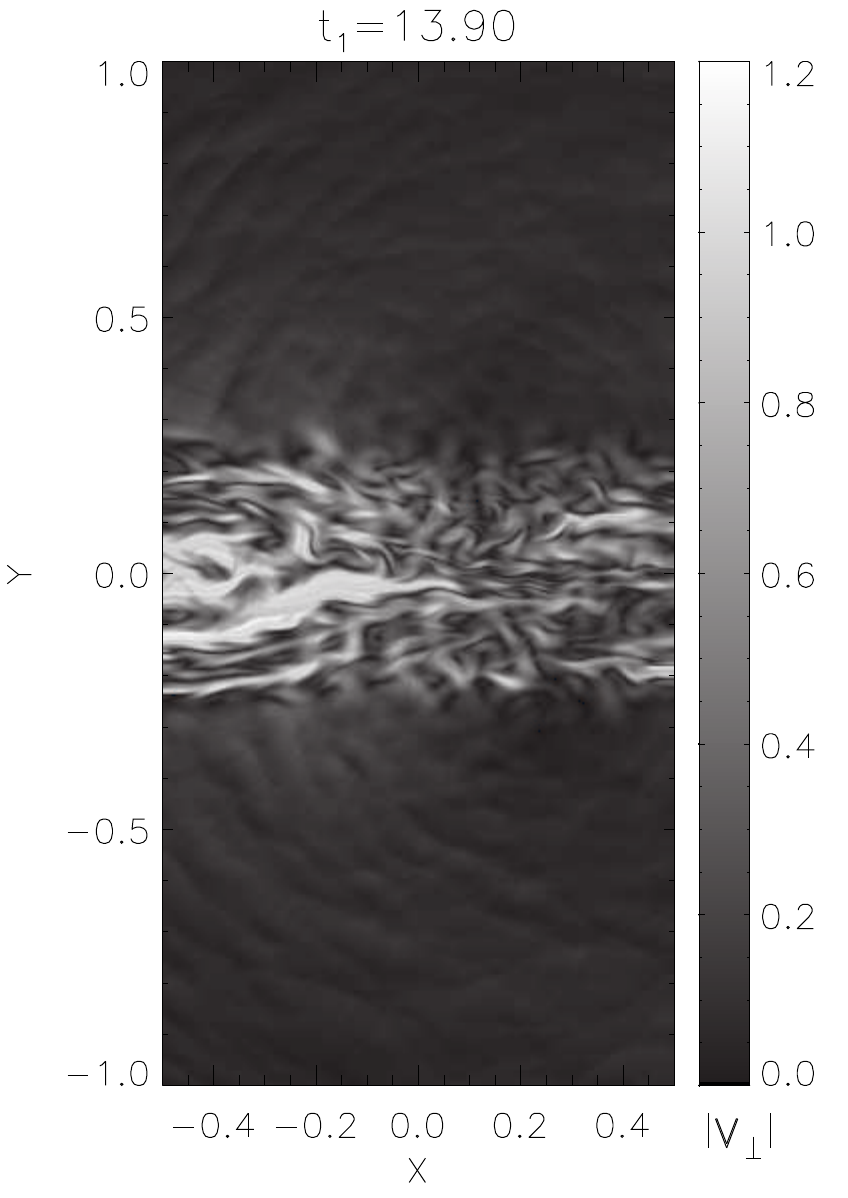}
\includegraphics[width=0.3\textwidth]{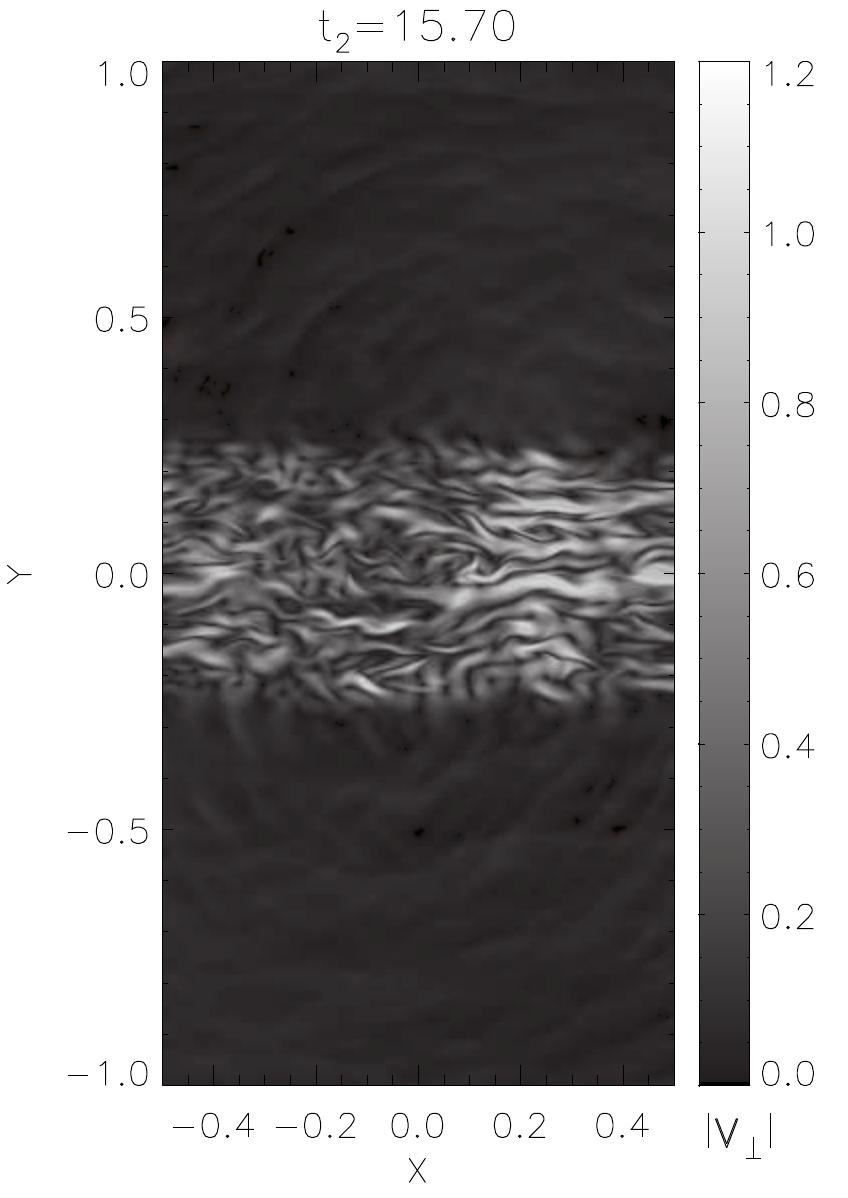}
\includegraphics[width=0.3\textwidth]{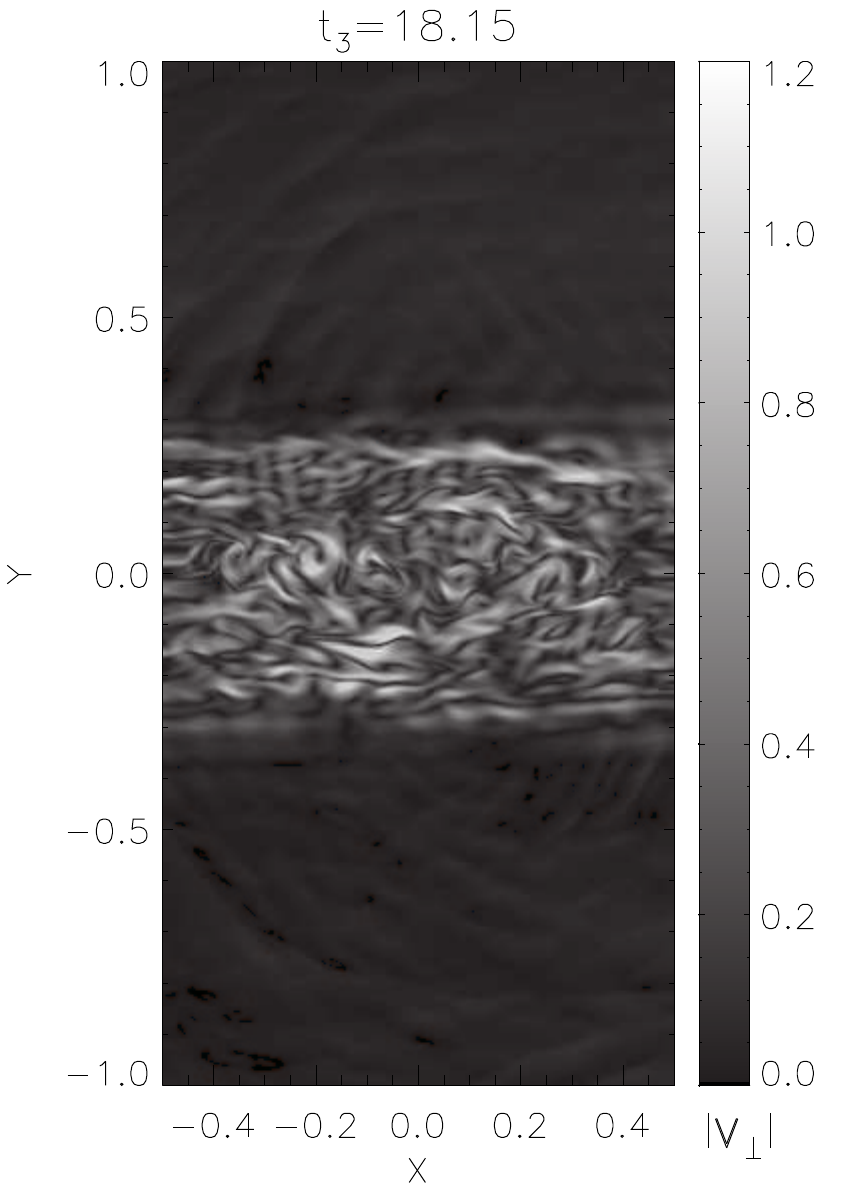} \\
\includegraphics[width=0.3\textwidth]{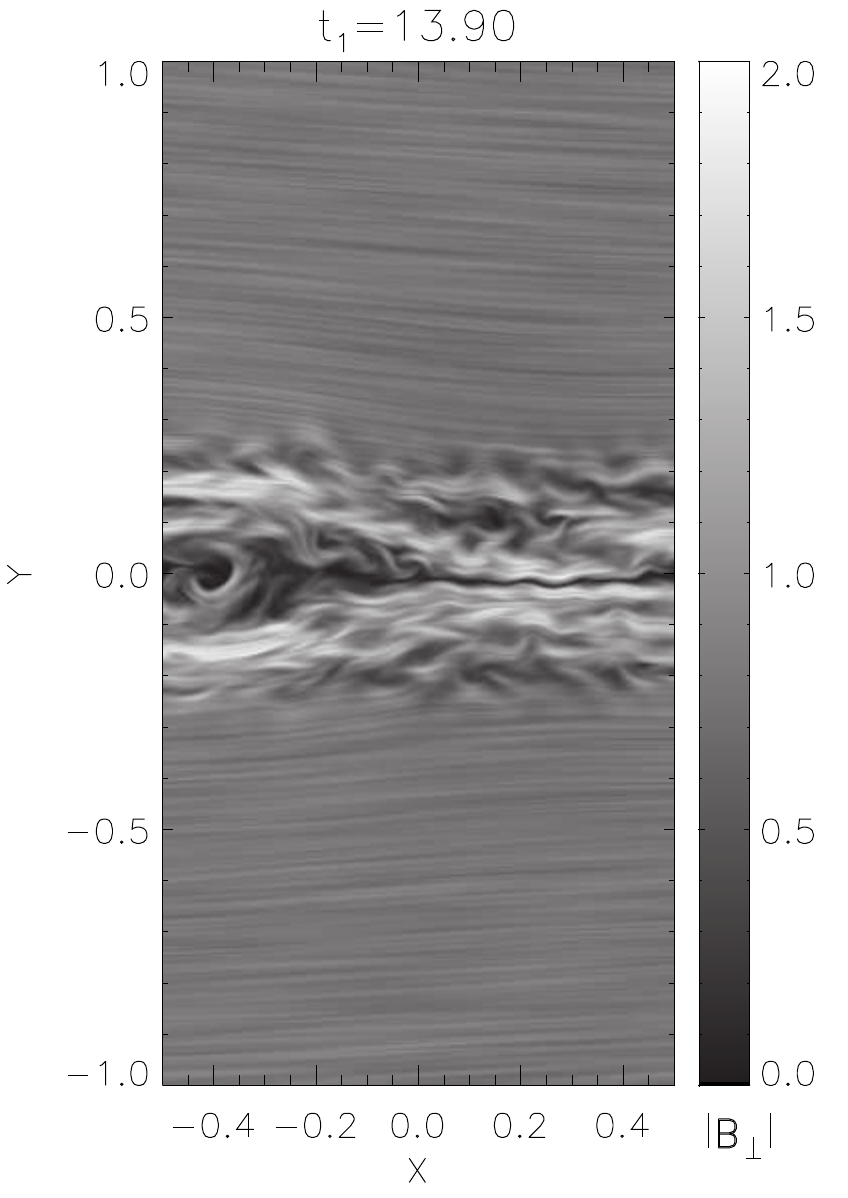}
\includegraphics[width=0.3\textwidth]{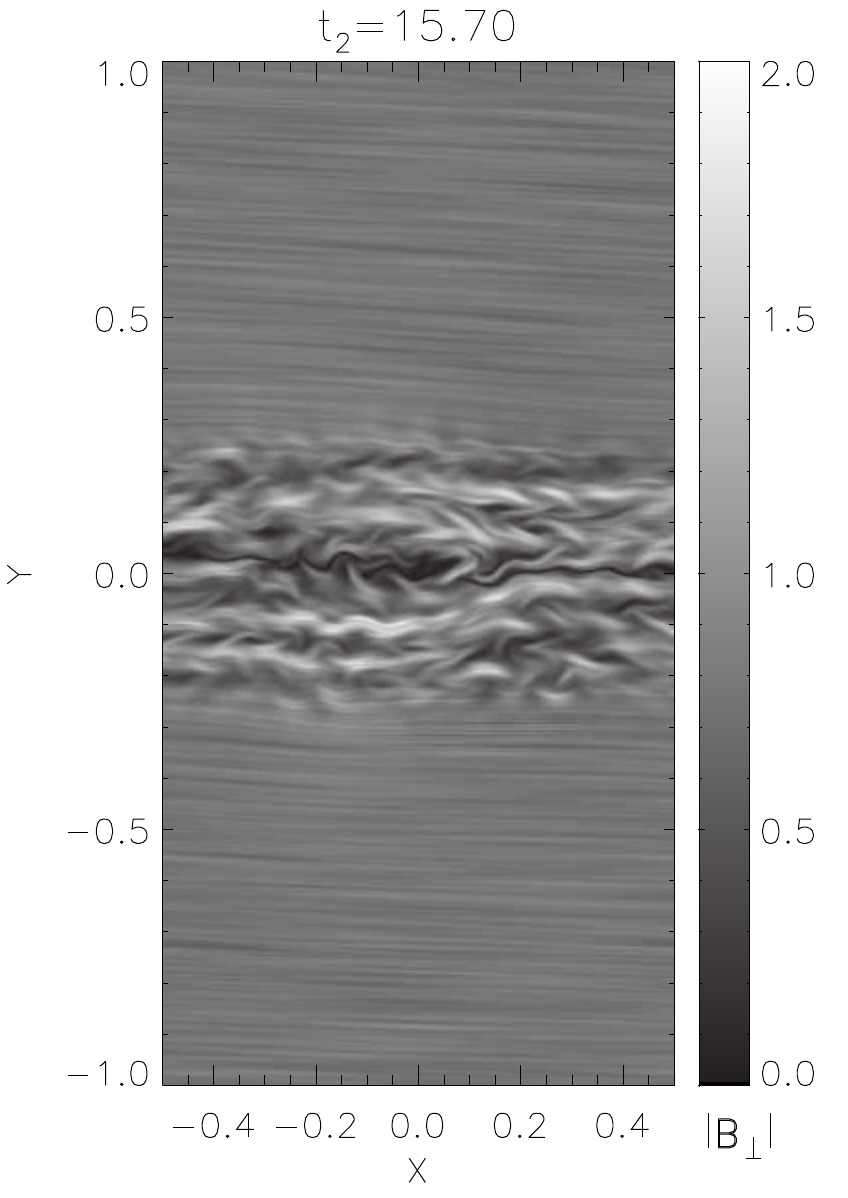}
\includegraphics[width=0.3\textwidth]{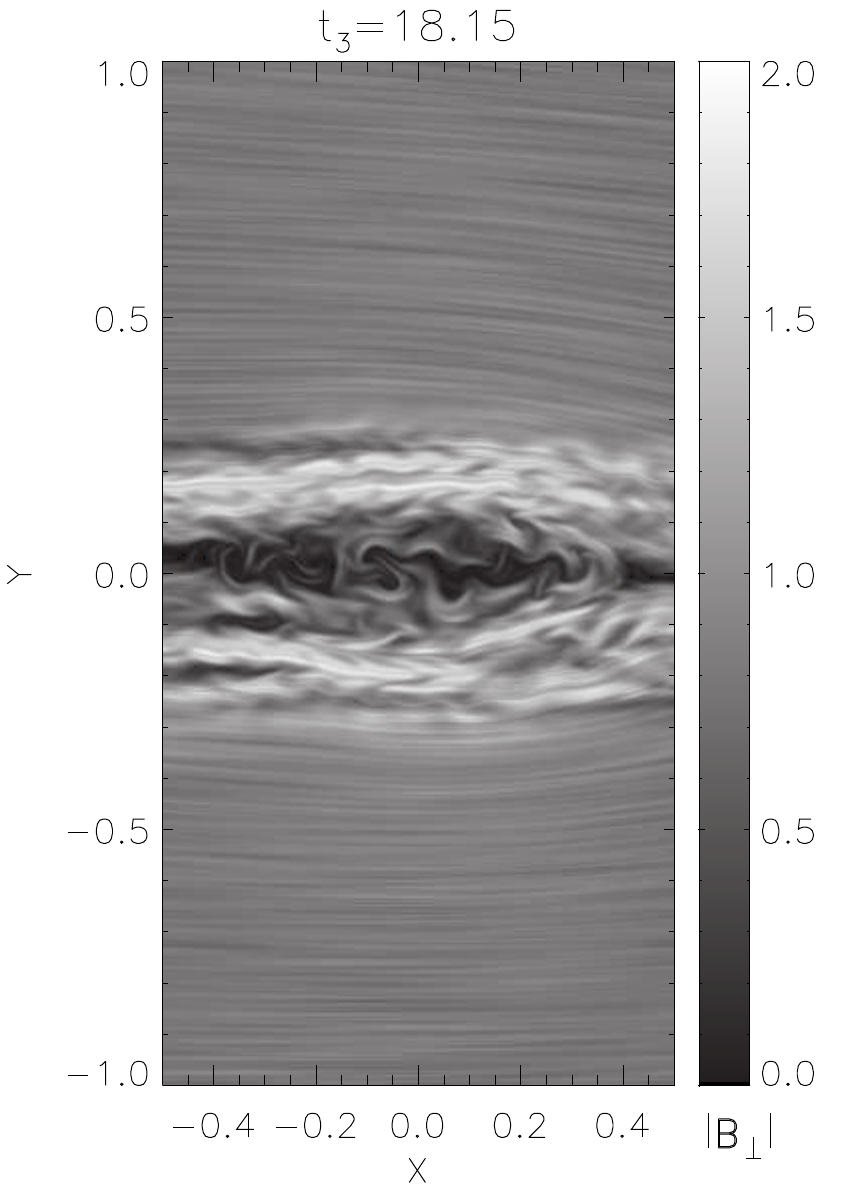} \\
\includegraphics[width=0.3\textwidth]{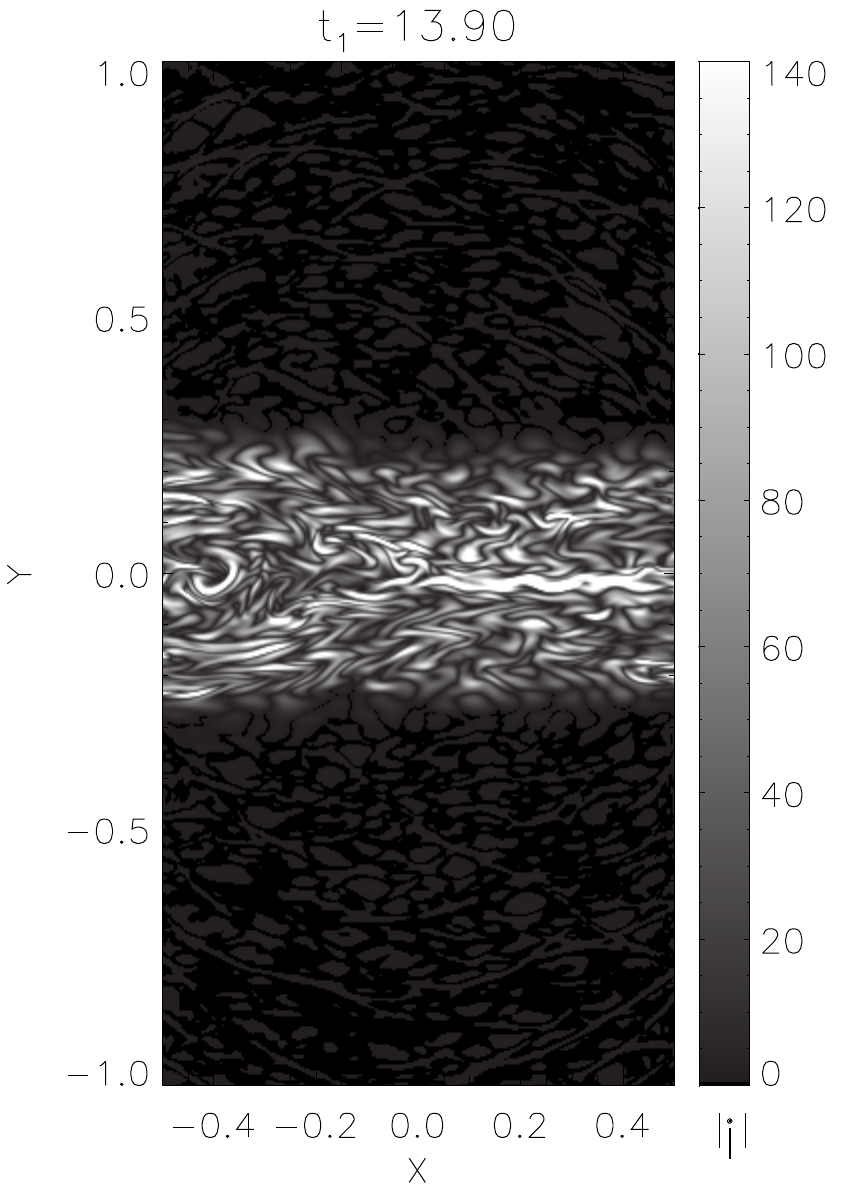}
\includegraphics[width=0.3\textwidth]{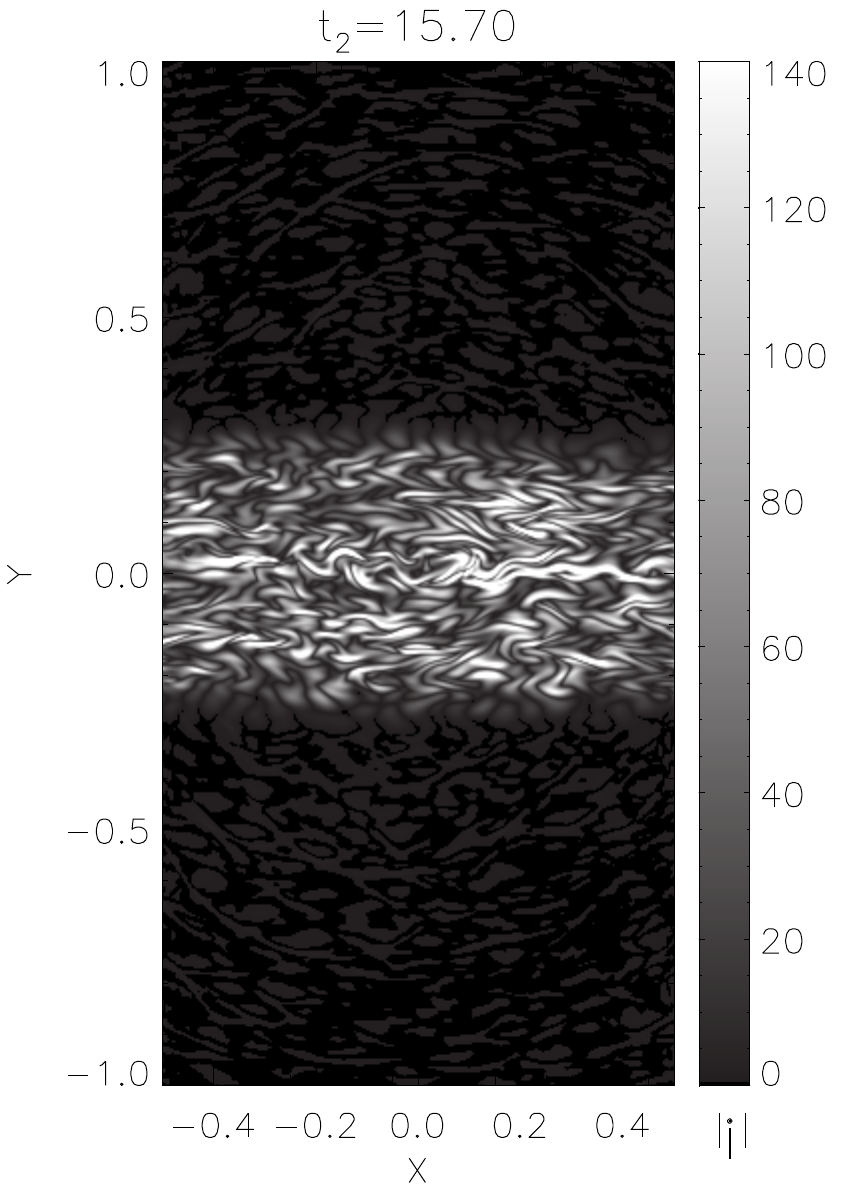}
\includegraphics[width=0.3\textwidth]{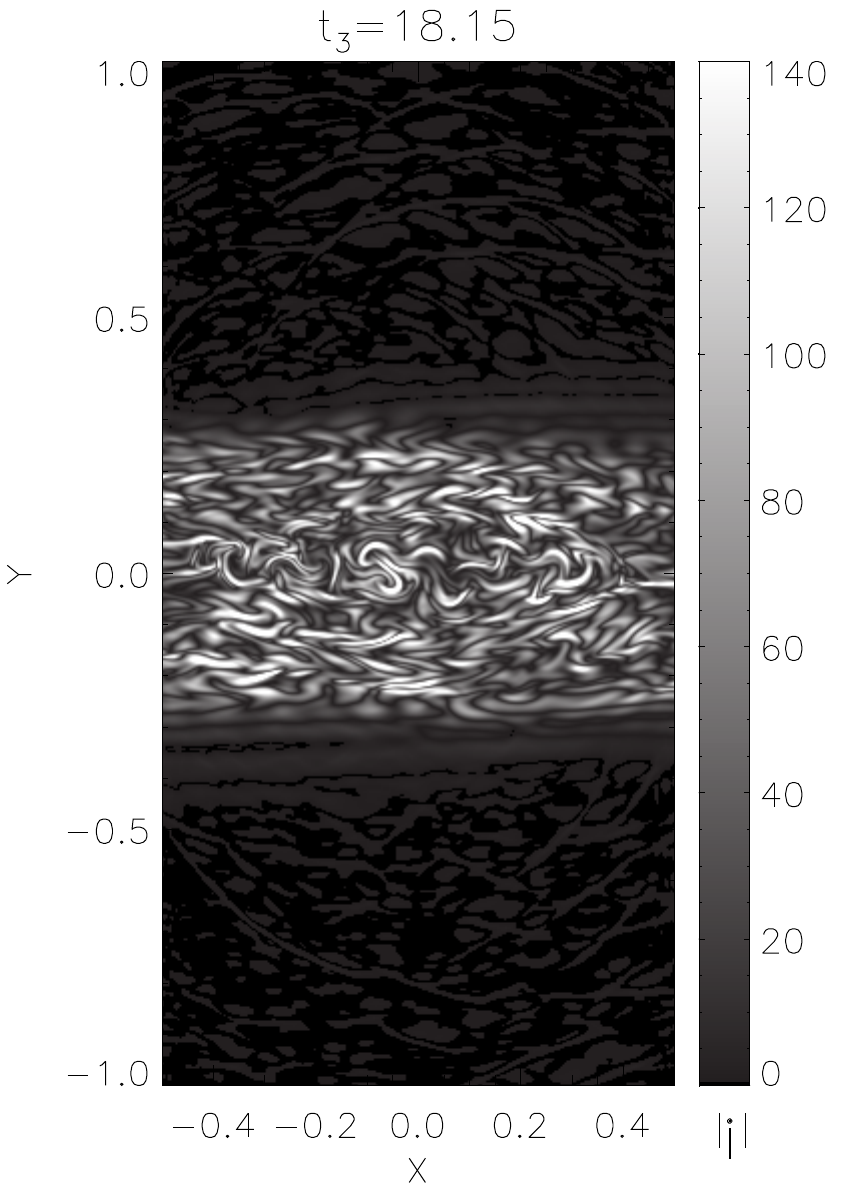}
\caption{Topology of velocity (top panels) and  magnetic field (middle panels)
in the presence of turbulence shown at three times $t_1=13.90$, $t_2=15.70$,
$t_3=18.15$.  In the lower panels we plot the absolute value of the current density
for the same times.  The evolution of the reconnection speed $V_r^{TB}$ for this
model is shown in Figure~\ref{fig:vr_p0.5}.  Turbulence is injected with  $P_{inj}=0.1$ at the wavenumber
$k=12$.  The uniform resistivity $\eta_u$ is equal $5\cdot10^{-4}$.  Brighter
shades correspond to larger values of displayed quantities. \label{fig:ev_p0.5}}
\end{figure*}

Figure~\ref{fig:vr_p0.5} shows the evolution of the reconnection speed for
$P_{inj}=0.5$, five times larger than in the
previous case. As before the reconnection rate does not reach a stationary
state. We observe a few maxima, however they do not appear periodically as in the
previous case (Figure~\ref{fig:vr_p0.1}). In fact, we have two main well defined
maxima at times $t\sim14.6$ and $t\sim19.5$.  The peak in the reconnection
speed is again caused by a fast outflow of gas visible at time 
$t_1=13.90$ in Figure~\ref{fig:ev_p0.5} (first column, top panel). If we look at
the  magnetic field topology at the same time (Figure~\ref{fig:ev_p0.5}
first column, middle panel) we see a magnetic island  which appears in the region 
of fast outflow.

At next time step ($t_2= 15.70$) the topology of magnetic and velocity fields are
still very mixed and complex (Figure~\ref{fig:ev_p0.5}, second column, top and
middle panel), however the magnetic field does not exhibit the very strong bending 
seen in the previous step. As the simulation proceeds the reconnection rate slowly
decreases, reaching even slightly negative values\footnote{This should not be taken too literally.  Our measure of the reconnection speed is misleading when the midplane magnetic field topology becomes progressively more tangled.} (Figure~\ref{fig:vr_p0.5}, from
$t\sim16$ to $t\sim18.5$).  

At time step $t_3=18.15$ (Figure~\ref{fig:ev_p0.5}, last column)  we show how our
model looks when the measured reconnection rate is negative. We can see a strong
accumulation of the magnetic field around the midplane which extends  over
almost the whole computational domain. The region of velocity fluctuations is broadened in
comparison to the previous time steps, even though the volume within which we drive turbulence is unchanged.   This quite unstable situation sets off
the extreme growth of reconnection rate (from $t\sim 18.5$ to $t\sim 19.5$)
shown in Figure~\ref{fig:vr_p0.5}. The current density distribution is plotted
in the bottom row in Figure~\ref{fig:ev_p0.5}. At every time step  many small
reconnection events occur, as indicated by local growth of the current
density.

\subsubsection{Dependence on the  Strength of Turbulence}

In order to check how the reconnection rate  $V_t^{TB}$ depends on the strength
of turbulence we make several simulations with different values of turbulent
power $P_{inj}$ (see model PD in Table~\ref{tab:models}). The rest of input
parameters have the same value in all these models.

\begin{figure}
\center
\includegraphics[width=\columnwidth]{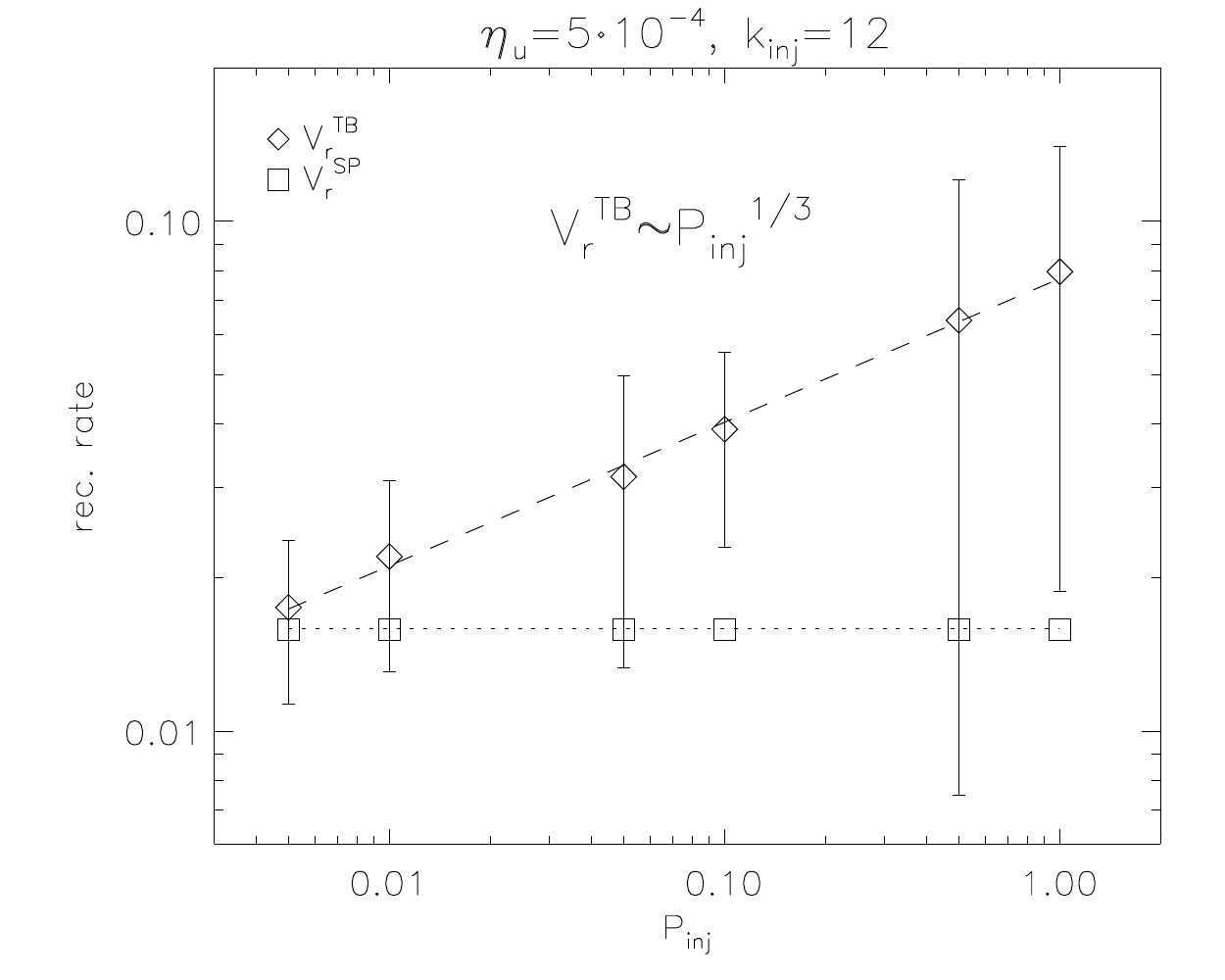}
\caption{Dependence of the reconnection rate on the power of injected turbulence
$P_{inj}$ for models with and without turbulence, $V_r^{TB}$ and $V_r^{SP}$,
respectively.  In all models we use the same values of the uniform resistivity
$\eta_u=5\cdot10^{-5}$ and the injection scale $k_{inj}=12$.  The error bars indicate the variance
of the reconnection rate.  In the case of
Sweet-Parker reconnection (squares) the errors are neglected and not shown.
\label{fig:vr_p}}
\end{figure}

In Figure~\ref{fig:vr_p} we present the dependence of the averaged reconnection
speed on the power of injected turbulence. Diamonds and squares correspond
respectively to the reconnection rate with and without turbulence. Both of them
are averaged over the fixed period of time: from $t=7$ to $t=9$ - laminar
reconnection (Sweet-Parker stage) and from $t=10$ to $t=20$ - turbulent
reconnection. The variance of the reconnection rate is calculated using the
standard deviation method. In case of the Sweet-Parker reconnection the variance
is negligible and is  not shown here. On the other hand, in the presence of turbulence the
reconnection rate undergoes continuous strong variations and the variance is only  slightly
smaller than the mean reconnection rate. We find that the reconnection rate
increases with power of turbulence and scales as $\sim P_{inj}^{1/3}$.

\subsubsection{Dependence on the Injection Scale}

Similarly we determine the dependence of the reconnection rate $V_r^{TB}$ on the
wavenumber at which we inject turbulence $k_f$. We compute several models with
varying   $k_f$ while keeping all other parameters the same (see
model PD in Table \ref{tab:models}).

\begin{figure}
\center
\includegraphics[width=\columnwidth]{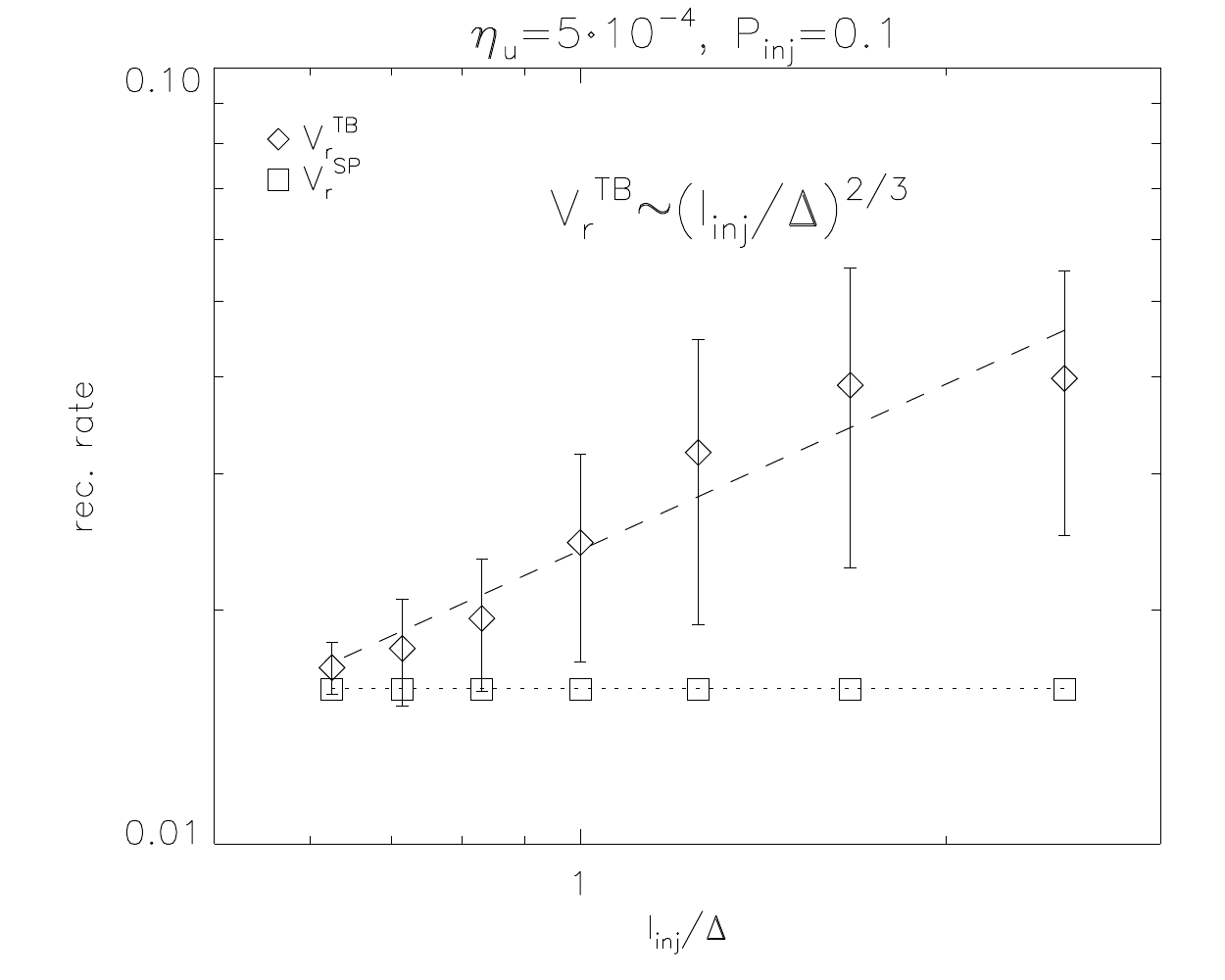}
\caption{Dependence of the reconnection rate on the injection scale $l_{inj}$
for models with and without turbulence, $V_r^{TB}$ and $V_r^{SP}$, respectively.
 In all models we use the same values of the uniform resistivity
$\eta_u=5\cdot10^{-5}$ and the power of injected turbulence $P=0.1$.  In the
case of Sweet-Parker reconnection (squares) the variance is negligible  and not
shown. \label{fig:vr_k}}
\end{figure}

Figure~\ref{fig:vr_k} shows the resulting dependence. We can clearly see the strong 
relation between injection scale $k_f$ and the reconnection speed. Namely, the reconnection rate
and errors increase with decreasing injection wavenumber $k_f$. Fitting to the simulation
results gives  $V_r^{TB} \sim l_{inj}^{2/3}$.

\subsubsection{Dependence on the Resistivity}

The most important goal in our study is to check the dependence of the
reconnection rate on the uniform resistivity. We run several
simulations with  different values of uniform resistivity $\eta_u$ and power of
turbulence $P_{inj}$ (see model RD in Table \ref{tab:models}).

\begin{figure}
\center
\includegraphics[width=\columnwidth]{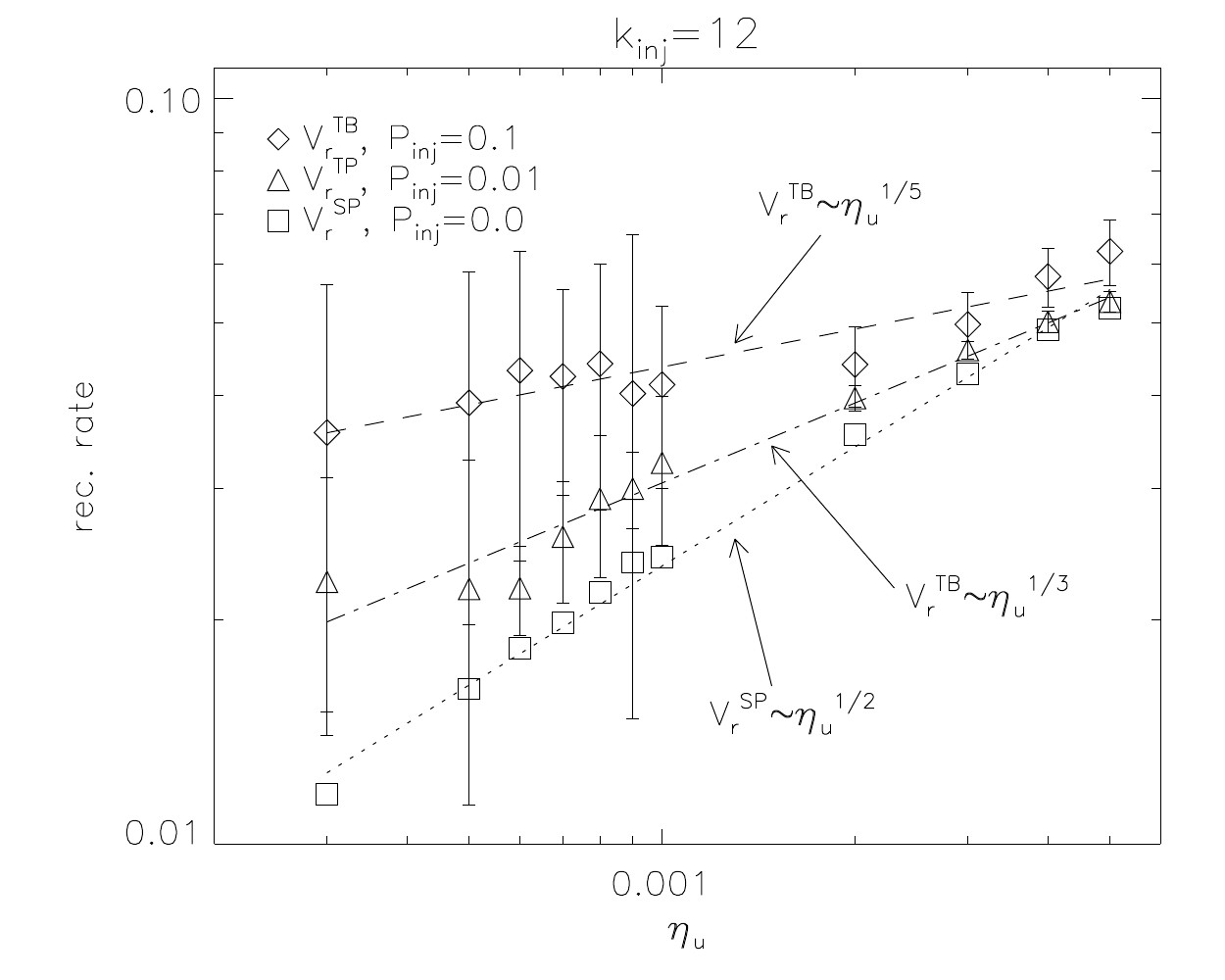}
\caption{Dependence of the reconnection rate on the uniform resistivity $\eta_u$
for models with and without turbulence, $V_r^{TB}$ (diamonds and triangles) and
$V_r^{SP}$ (squares), respectively.  Turbulence is injected at the scale $k=12$
with $P_{inj}=0.1$ and $P_{inj}=0.01$.  In the case of Sweet-Parker reconnection
(squares) the variance is negligible and not shown. \label{fig:vr_r}}
\end{figure}

In Figure~\ref{fig:vr_r} we plot the obtained dependence of reconnection rate on
uniform resistivity. For Sweet-Parker configuration the averaged reconnection
rate $V_r^{SP}$ is represented by squares. Dispersion of calculated points is
almost negligible and not shown. As we can see the reconnection speed
$V_r^{SP}$ increases with uniform resistivity. From the fitting we obtain that
 the reconnection rate $V_r^{SP}$ scales with the uniform resistivity as
$\sim \eta_u^{1/2}$.  This is in agreement with theoretical prediction
\citep{swe-58,par-57} and confirms that our 2D model works well in the laminar
reconnection stage.

Adding turbulence to the system leads to a weaker  dependence between the reconnection
rate and  the uniform resistivity. Moreover,
the power of turbulence also influences  this relationship, as is clearly
visible in Figure~\ref{fig:vr_r} (triangles - $P_{inj}=0.01$ , squares -
$P_{inj}=0.1$). We fit lines to calculated points and find that the dependence
between reconnection rate and uniform resistivity is stronger for lower values
of $P_{inj}$. Namely, for $P_{inj}=0.1$ and $P_{inj}=0.01$  we get that the
reconnection rate scales as $\sim \eta_u^{1/5}$ and $\sim \eta_u^{1/3}$,
respectively.

For low values of the uniform resistivity ($\eta_u\leq1\cdot10^{-3}$-$P_{inj}=0.1$ and
$\eta_u\leq7\cdot10^{-4}$-$P_{inj}=0.01$) we see that the obtained reconnection
speeds $V_r^{TB}$  are almost the same.  This similar values of reconnection
rate  may be caused by prevailing numerical diffusion. Then, taking into account
higher values of $\eta_u$ does not modify the overall value of the total
resistivity. However, in this case the numerical diffusion will also influence
the reconnection rate in the Sweet-Parker configuration, what does not happen in
our simulations (see Figure~\ref{fig:vr_r}).

Results described above indicate that in 2D turbulent reconnection depends on
the uniform resistivity.  However, the obtained dependencies are not certain
because of the large dispersion of results for small values of the uniform
resistivity $\eta_u$.

\begin{figure}
\center
\includegraphics[width=\columnwidth]{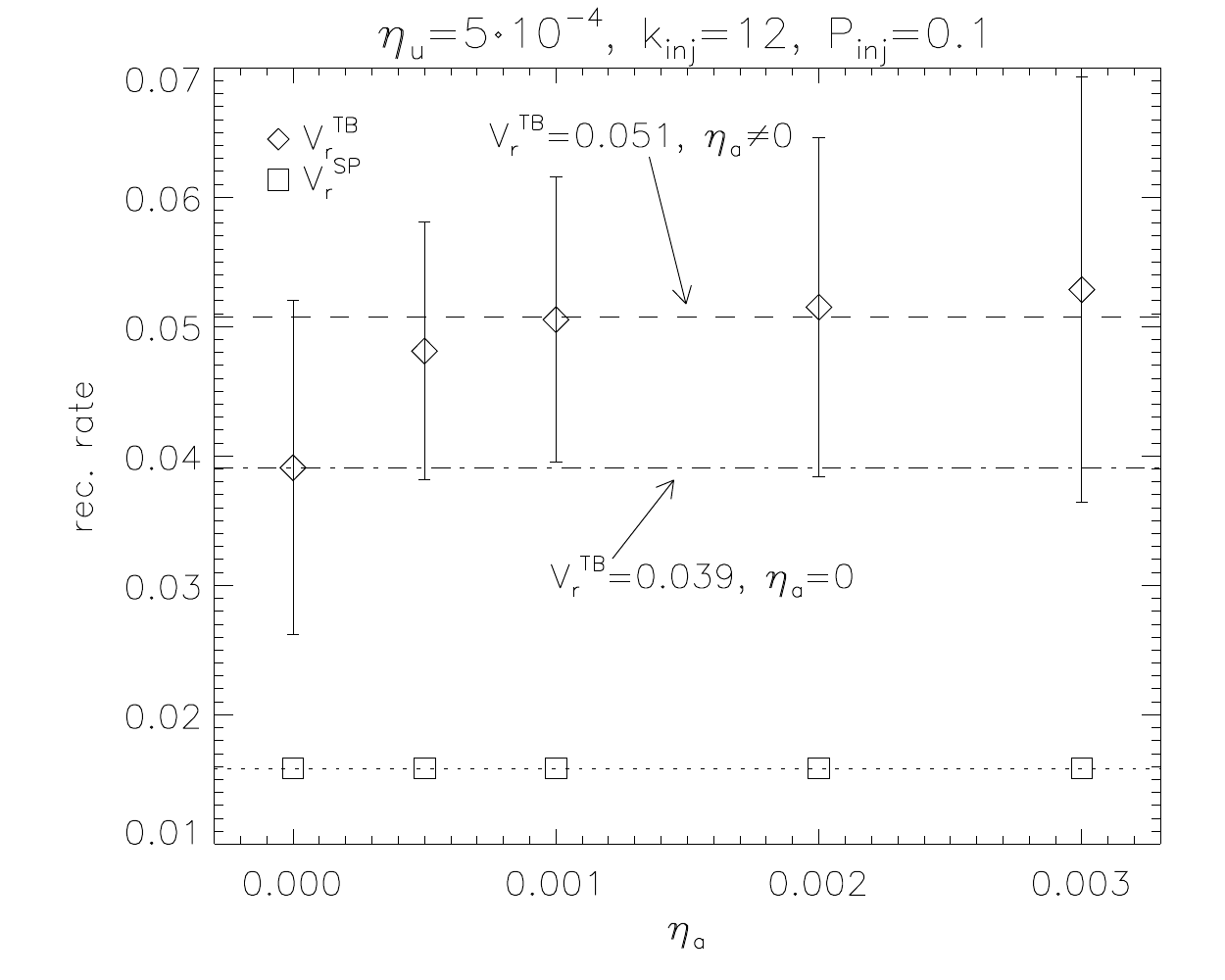}
\caption{Dependence of the reconnection rate on the anomalous resistivity
$\eta_a$ for models with and without turbulence, $V_r^{TB}$ (diamonds) and
$V_r^{SP}$ (squares), respectively.  Turbulence is injected at the wavenumber
$k_f=12$ with $P_{inj}=0.1$.  The uniform resistivity $\eta_u$ is
equal $5\cdot10^{-4}$.  The dotted line represents the mean reconnection rate
obtained for the Sweet-Parker stage, dash-dotted marks the value of the
reconnection rate without anomalous resistivity and dashed line corresponds to
the mean reconnection rate calculated for four models with different values of
the anomalous resistivity. \label{fig:vr_a}}
\end{figure}

We also test the dependence of the reconnection rate on the anomalous
resistivity. We run several models with the same value of uniform resistivity
$\eta_u=5\cdot10^{-4}$ and critical current density $j_{crit}=50$ but for
different values of the anomalous resistivity $\eta_a$ (see model AD in Table
\ref{tab:models}).  In Figure~\ref{fig:vr_a} we plot the reconnection rate
calculated for the Sweet-Parker configuration (squares) and in the presence of  turbulence
 (diamonds). The dash-dotted line determines the mean value of the
reconnection rate ($\overline{V_r^{TB}}_{(\eta_a\neq0)}=0.051$) obtained for
models with the anomalous resistivity. As we see, the presence of
anomalous resistivity causes an increase in the reconnection speed compared
to a model with $\eta_a=0$ - dotted line ($V_r^{TB}=0.039$).  However, for
different values of the anomalous resistivity the reconnection speed is almost
the same. More precisely, we observe a nearly negligible increase in the reconnection
rate and a clear increase in the variance at  larger values of
the anomalous resistivity $\eta_a$.


\subsection{Dependence on the Viscosity}
\begin{figure}
\center
\includegraphics[width=\columnwidth]{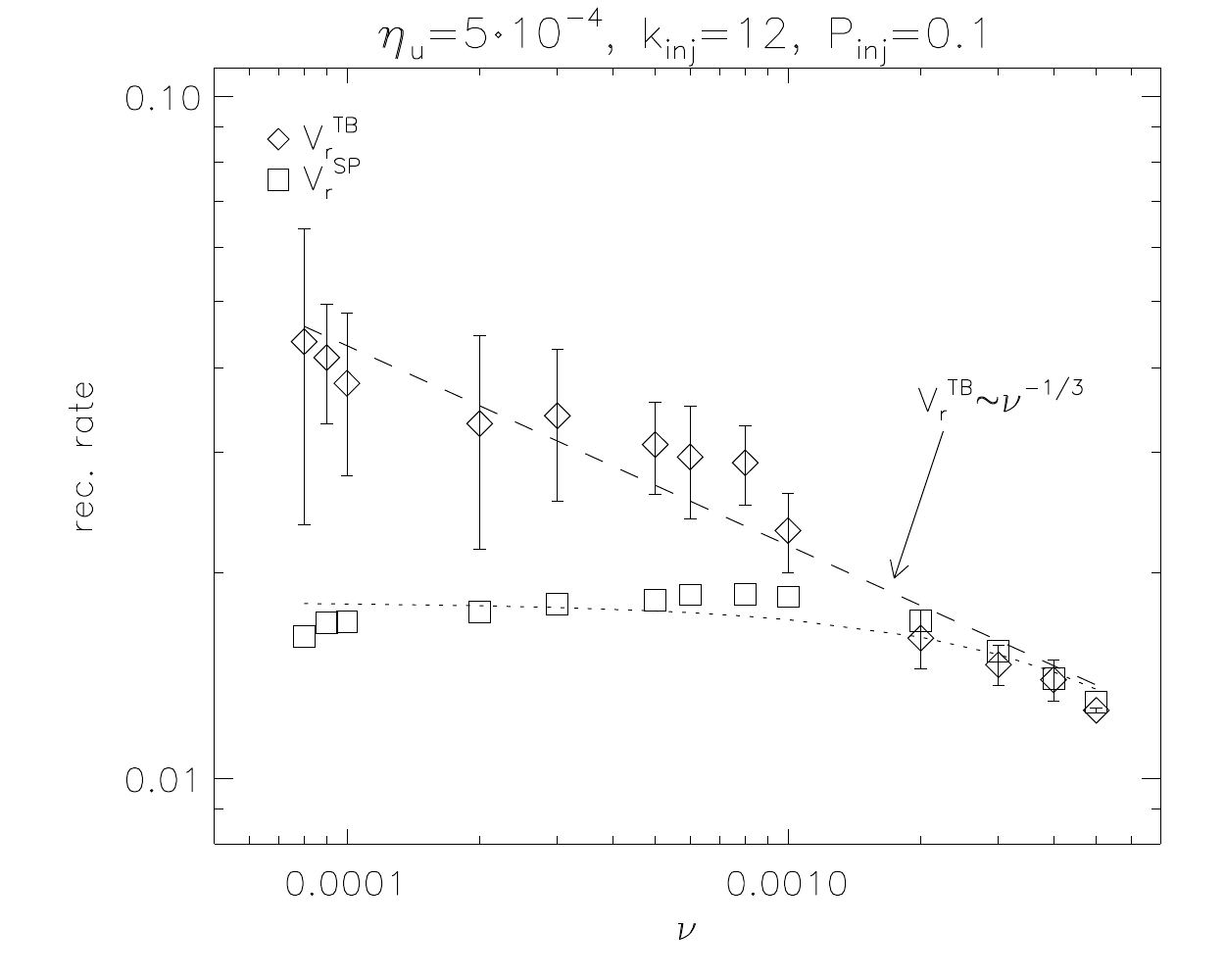}
\caption{Dependence of the reconnection rate on viscosity $\nu$ for models with
and without turbulence, $V_r^{TB}$ and $V_r^{SP}$, respectively.  In all models
we use the same values of the uniform resistivity $\eta_u=5\cdot10^{-5}$,
turbulence strength $P=0.1$ and the injection wavenumber $k_{inj}=12$.  In the case
of Sweet-Parker reconnection (squares) the variance is negligible and not shown.
 \label{fig:visc}}
\end{figure}

Next we test the influence of viscosity on the reconnection rate for fixed
values of the injection scale, turbulence strength and uniform resistivity
(see model VD in Table~\ref{tab:models}). The results are shown in
Figure~\ref{fig:visc} where we plot averaged reconnection rates for the laminar
(squares) and turbulent (diamonds) cases. In the Sweet-Parker configuration the
reconnection rate $V_r^{SP}$ is almost the same for viscosity
$\nu\leq1\cdot10^{-3}$ and starts to decrease for $\nu\geq2\cdot10^{-3}$. In the
present of turbulence the fit to the calculated points shows that the
reconnection rate $V_r^{TB}$ scales with the viscosity as $\sim \nu^{-1/3}$.
Furthermore, when the viscosity is larger than $1\cdot10^{-3}$ the rate of
reconnection for models with and without turbulence are similar.

\subsection{The Case of a Uniform Initial Configuration}

In order to check the influence of our boundary conditions and method
of driving turbulence on the reconnection process we compare two models with
different initial magnetic field configurations: antiparallel and uniform. The
antiparallel initial configuration of the magnetic field is the same as in the
case of the Sweet-Parker reconnection,  the uniform one is horizontal with
$B_x=1$. The rest of parameters characterizing both models are:
$\eta_u=5\cdot10^{-4}$, $k_f=12$, $P=0.1$ and inflow/outflow boundary
conditions. In both models we inject turbulence in the same way from $t=9$ to
$t=10$.

\begin{figure}
\center
\includegraphics[width=\columnwidth]{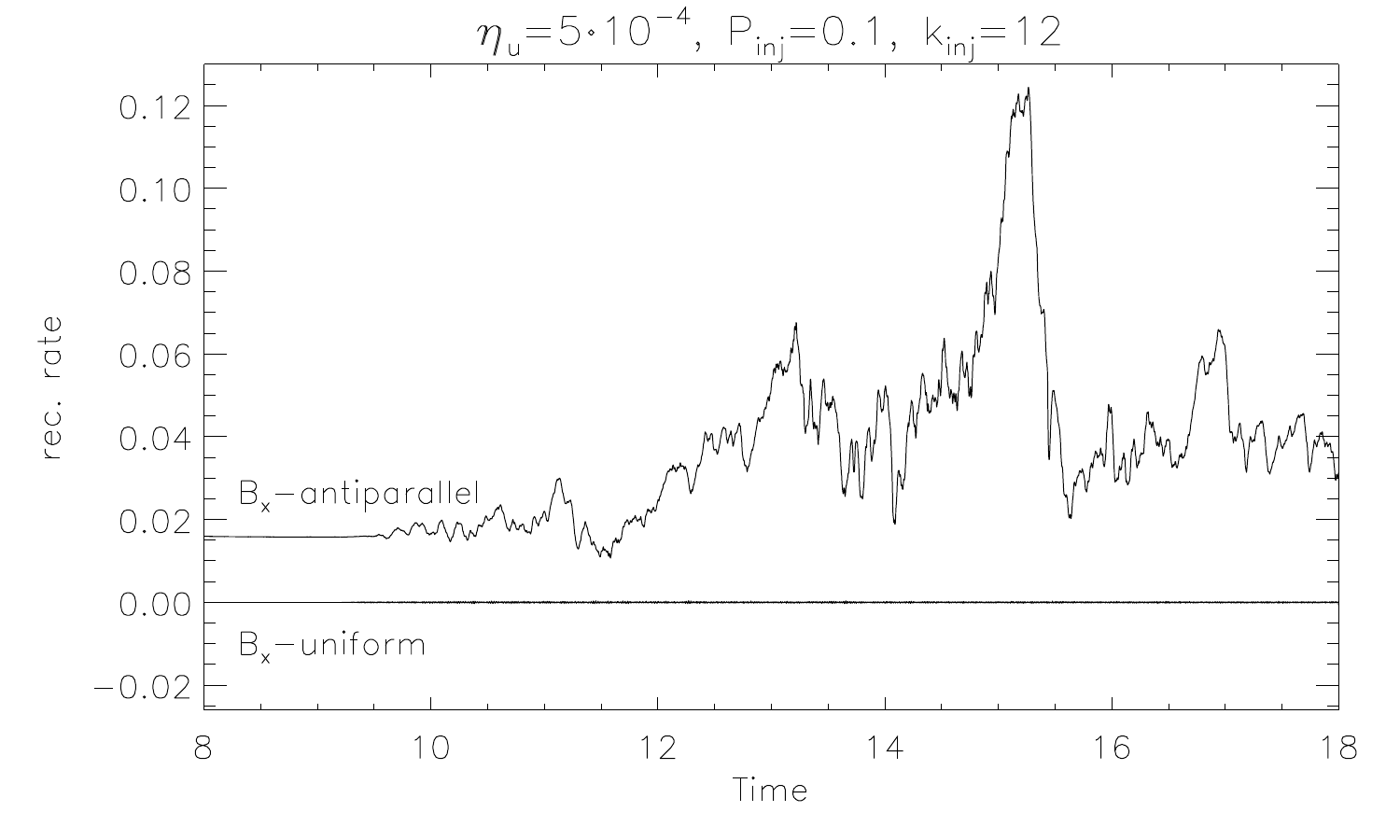}
\caption{Comparison of the reconnection rates obtained for models with
uniform and antiparallel initial magnetic field configurations.  All parameters
describing models are the same: the power of turbulence $P_{inj}=0.1$, the
injection wavenumber $k_f=12$, and the uniform resistivity coefficient
$\eta_u=5\cdot10^{-4}$.  The topology of the velocity and magnetic velocity fields,
as well as the distribution of the current density for both models are shown in
Figure~\ref{fig:ev_p0.1} (antiparallel initial magnetic field) and in
Figure~\ref{fig:turb} (uniform initial magnetic field). \label{fig:c_vt}}
\end{figure}

In Figure~\ref{fig:c_vt} we show the comparison of reconnection rates obtained for
the uniform and antiparallel initial magnetic field configuration. In the first
case the reconnection rate is almost equal zero. We can see some small
fluctuations, however we cannot observe any significant growth of the
reconnection rate. In the second case the value of reconnection rate is higher
and changes violently. For more precisely description of this case see Sect.
\ref{turb_reconn}.

\begin{figure*}
\center
\includegraphics[width=0.3\textwidth]{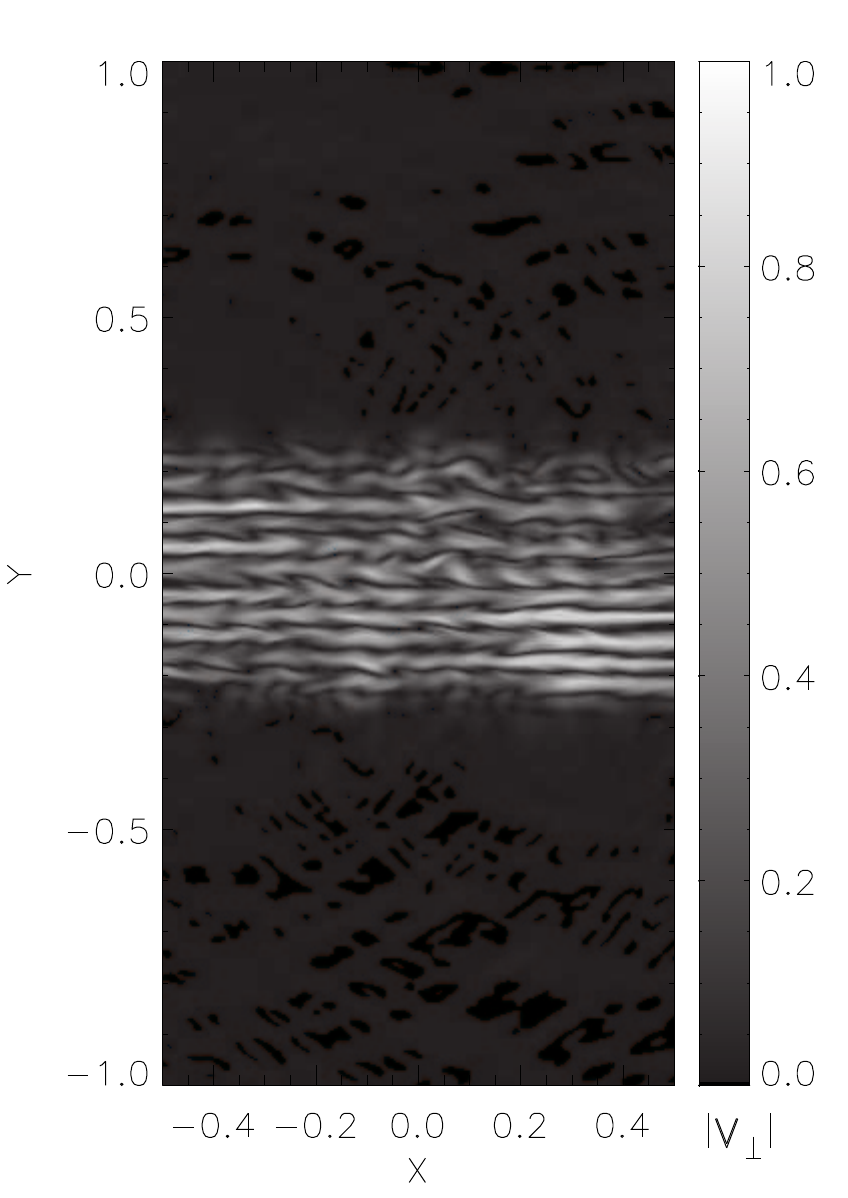}
\includegraphics[width=0.3\textwidth]{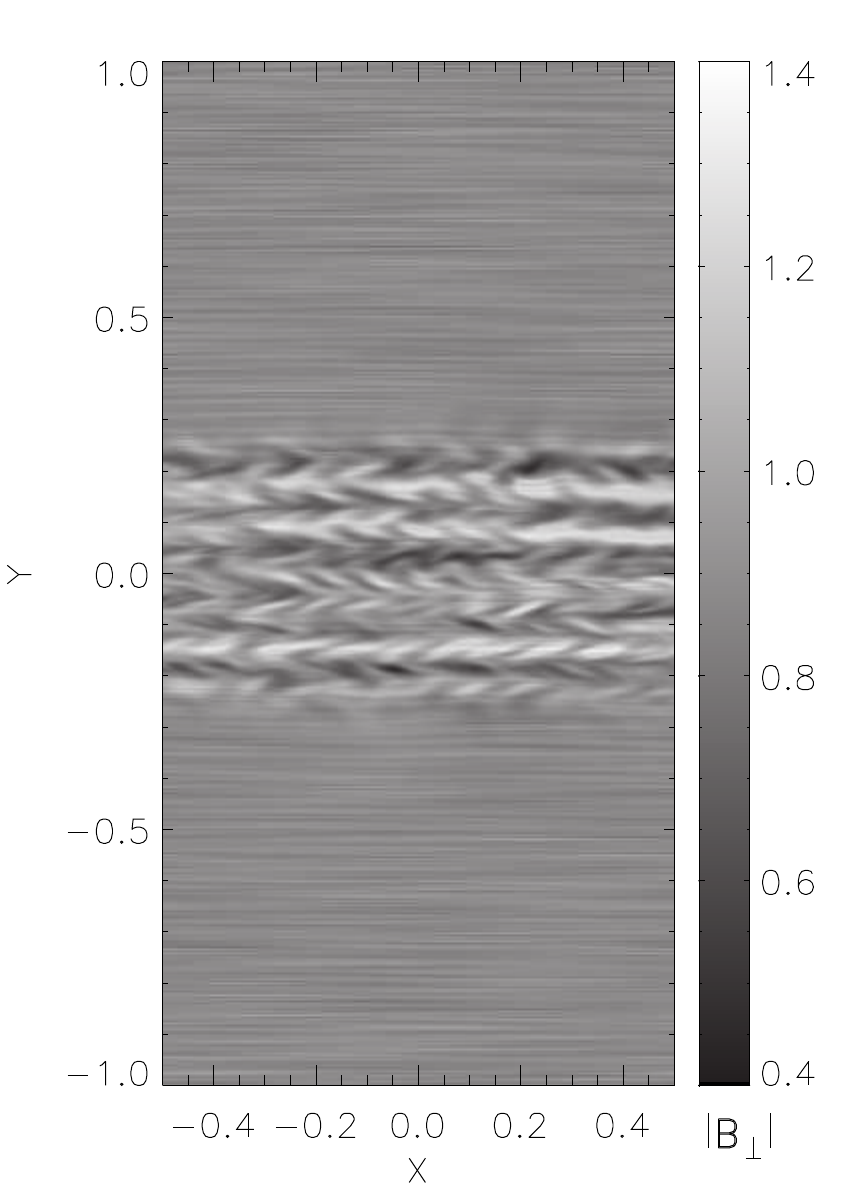}
\includegraphics[width=0.3\textwidth]{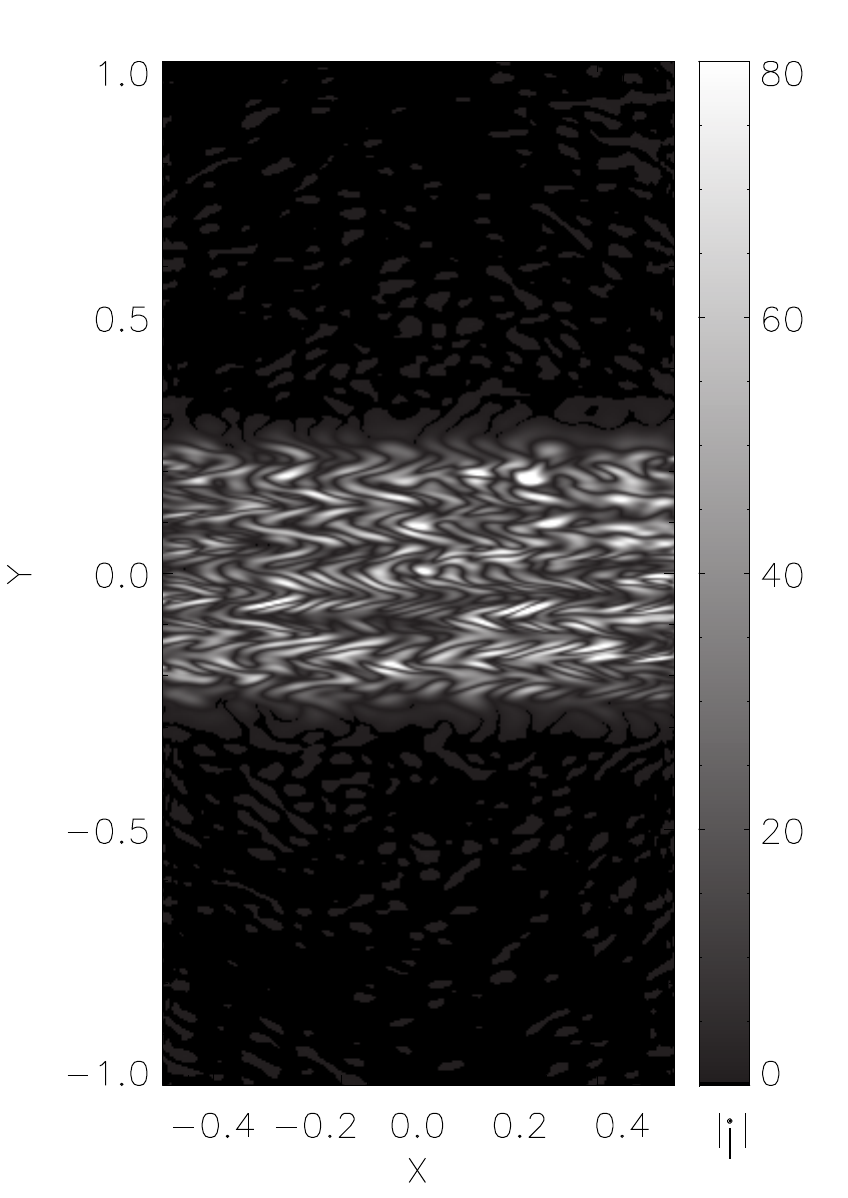}
\caption{Topology of  velocity (left panel) and magnetic field (middle panel),
and the absolute value of the current density distribution (right panel) in the
case of the initially uniform magnetic field at the time $t=14$.  Brighter
shades correspond to larger values of the displayed quantities.  Turbulence is
injected with the power $P_{inj}=0.1$ at wavenumber $k_f=12$.  The uniform
resistivity $\eta_u$ is equal $5\cdot10^{-4}$.  For comparison, the model with
the same set of parameters but with an initially antiparallel configuration of magnetic
field is presented in Figure~\ref{fig:ev_p0.1}. \label{fig:turb}}
\end{figure*}

The almost negligible value of the reconnection rate for the model with a uniform
initial magnetic field indicates that, as expected, magnetic reconnection does not occur.
This also supports the conclusion that our calculation of the reconnection rate is not an artifact
of our boundary conditions or our method for implementing turbulence. 

Figure~\ref{fig:turb} shows the topology of magnetic and velocity field as well as
the distribution of current density for model with initial uniform configuration
of the  magnetic field. As before turbulence is injected in a volume surrounding the
midplane which causes  perturbations of the velocity field in this region (see
the left panel in Figure~\ref{fig:turb}). Most of fluctuations propagate along the mean
magnetic field running more or less  parallel to the midplane  (see the middle panel
in Figure~\ref{fig:turb}). The well defined bend of magnetic lines presented in
the case of turbulent reconnection are not visible here. This is because we do
not have a diffusion region, where the magnetic lines change their direction
and strength, and can be bent by fluctuations. On the other hand, turbulence
causes small perturbations of magnetic field which leads to an enhancement of the
absolute value of the current density (see the middle panel in Figure~\ref{fig:turb}).

\subsection{Long Term Variance of the Reconnection Rate}

\begin{figure*}
\center
\includegraphics[width=\textwidth]{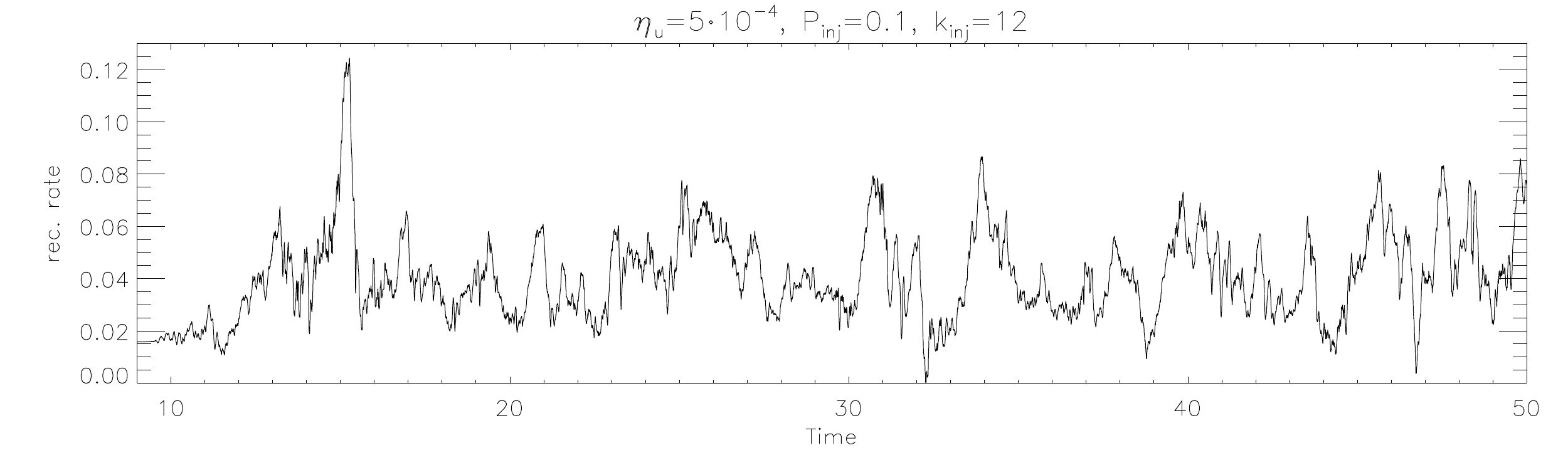}
\caption{Time evolution of the reconnection speed for the model with t $P_{inj}=0.1$, 
an injection wavenumber $k_f=12$ and a
uniform resistivity  $\eta_u=5\cdot10^{-4}$. \label{fig:vr_long}}
\end{figure*}

\begin{figure}
\center
\includegraphics[width=0.6\columnwidth]{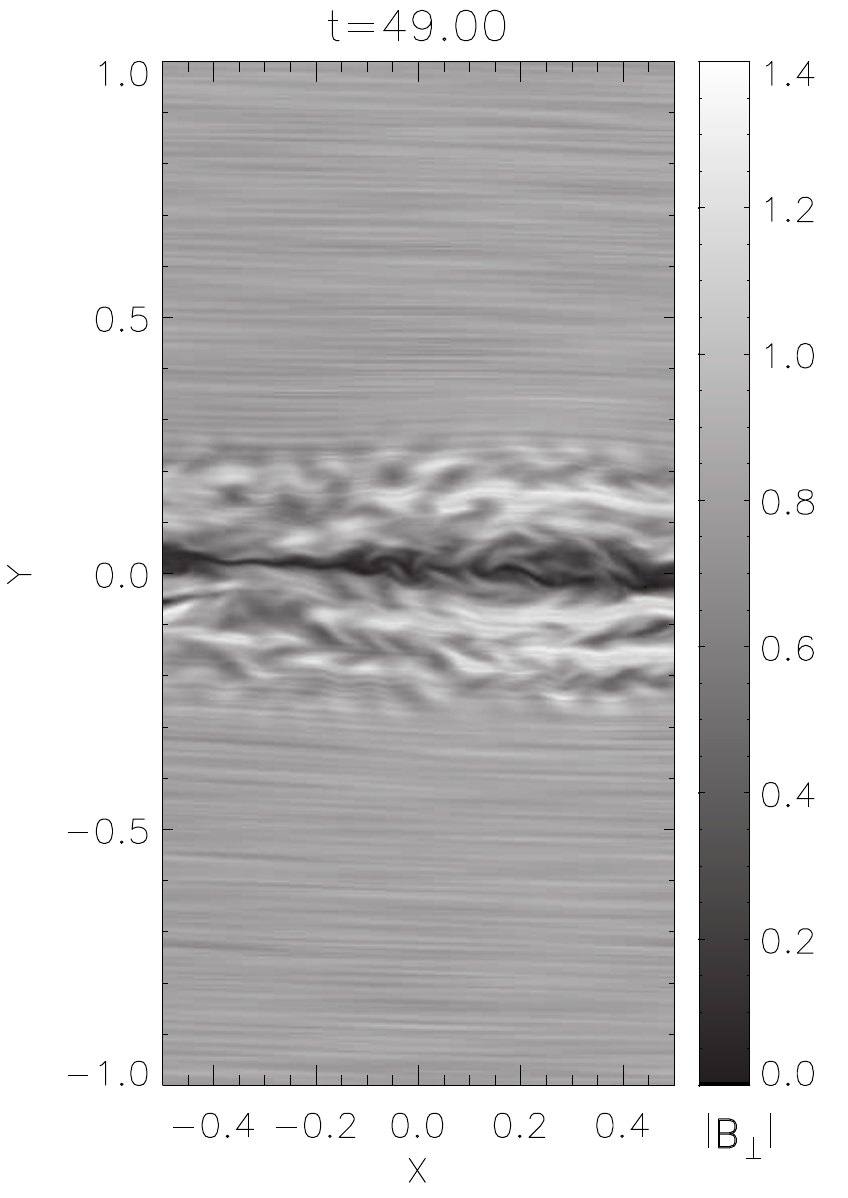}
\caption{Topology of magnetic field in the presence of turbulence at time
$t=49$.  The uniform resistivity $\eta_u$ is equal $5\cdot10^{-4}$.  Turbulence
is injected with the power $P_{inj}=0.1$ at scale $k=12$.  Brighter shades
correspond to larger values of the displayed quantities. \label{fig:btop_l}}
\end{figure}

The time evolution of the reconnection rate for the same model as in Figure~\ref{fig:vr_p0.1} 
but calculated over a longer period of time ($t_{end}=50$) is
shown in Figure~\ref{fig:vr_long}. We see that even after 50 Alf\'enic
times the reconnection rate does not reach a steady state. Strong and
continuous fluctuations are seen over the whole simulation time. All these quasiperiodic
changes of the reconnection rate are apparently driven by something like the  tearing mode
instability. Namely, big loops of magnetic field are continuously created (see
Figure~\ref{fig:btop_l}) and ejected through boundaries.

\begin{figure}
\center
\includegraphics[width=\columnwidth]{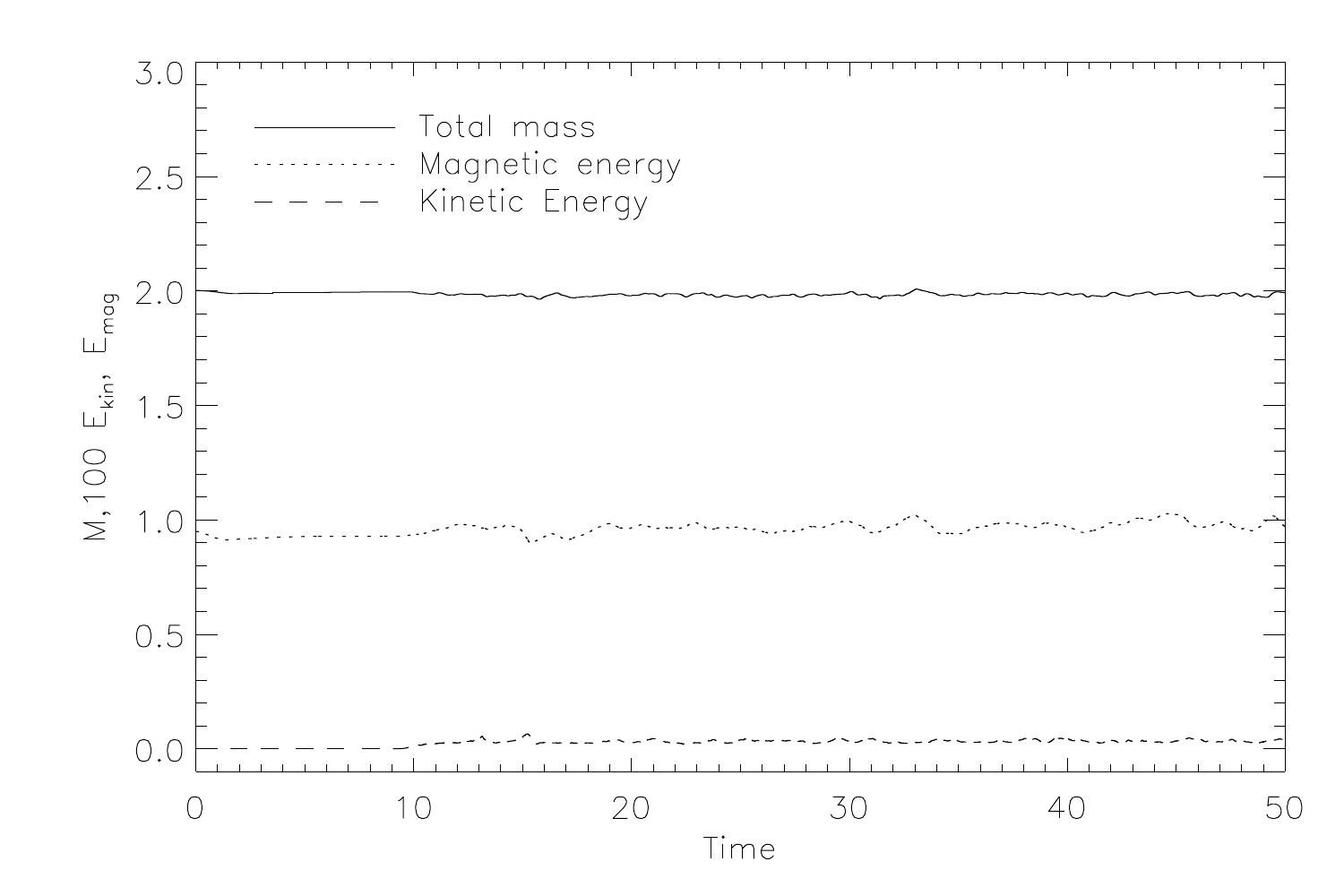}
\caption{Time evolution of the total mass $M$ (solid line), magnetic $E_{mag}$
(dashed line) and kinetic $E_{kin}$ (dotted line) energies during 
Sweet-Parker (from $t=0$ to $t=9$) and turbulent (from $t=9$ to $t=50$
reconnection with uniform resistivity $\eta_u=5\cdot10^{-4}$.  The kinetic
energy $E_{kin}$ is not amplified in this plot. \label{fig:en_long}}
\end{figure}

In Figure~\ref{fig:en_long} we show the evolution of mass, magnetic and kinetic
energy during the Sweet-Parker (from $t=0$ to $t=9$) and turbulent (from $t=9$
to $t=50$) reconnection. We see that adding turbulence  to the system or using
open boundary conditions does not introduce instabilities in the total mass or kinetic
and magnetic energies.

What is more, increasing $t_{end}$ from $20$ to $50$ only slightly influences  the
average reconnection rate and its variance. For models with $t_{end}=20$ and
$t_{end}=50$ we get $V_r^{TB}=0.039\pm0.019$ and $V_r^{TB}=0.041\pm0.017$,
respectively.

This indicates that extending the simulation time does not lead to a stable state
of reconnection and does not change our results.

\section{Discussion}
\label{discussion}

\subsection{The Goals of this Study}

The present paper provides a follow-up study to \cite{kow-09}.  In that paper fast
magnetic reconnection of a 3D weakly stochastic magnetic field, predicted in the
LV99 theoretical study, was confirmed.  Can turbulent reconnection be fast in 2D
as well?  The question may sound rather scholastic, as the nature we deal with
is definitely three dimensional.  However, there are at least two reasons why
answering this question is important.  First of all, the claim in LV99 is that
3D effects are essential for fast reconnection.  Thus it is
important to test this claim and to explore whether  fast reconnection in the
presence of turbulence can be carried over to 2D.  Then, in a number of earlier
studies it was conjectured that magnetic reconnection could become fast in the
presence of turbulence for purely  2D reconnection configurations.

In all respects apart from dimensionality the present
numerical set-up is identical to that in \cite{kow-09}.  In particular, the
excitation of turbulence in both case at subAlfv\'enic velocities, thereby preventing
field reversals arising from turbulence.  For measurements of the
reconnection rate we adopt the procedure presented in \cite{kow-09}.  This
enables us to make easy comparison between the 3D and 2D reconnection results.


\subsection{2D versus 3D Reconnection}

Until very recently 2D geometry has been favored  for magnetic
reconnection studies.  Its advantage stems from the fact that it allows one to 
achieve much higher Lundquist numbers compared to its 3D counterpart.  The most
important question is whether the nature of reconnection is the same in 3D and
in 2D.  It is suggestive that the answer to this question is positive if the
reconnection of laminar magnetic field is involved.  Our results on the
Sweet-Parker reconnection above are very similar\footnote{The obtained
dependence of reconnection pace on the uniform resistivity coincide in 2D and 3D
and is consistent with theory, i.e. $V_r^{SP}\sim\eta_u^{1/2}$.} to the results
of the Sweet-Parker reconnection in \cite{kow-09}.  At the same time our
comparison of the results in the present paper with those in \cite{kow-09} shows
that the reconnection in the presence of turbulence is very different.

The aforementioned difference is not surprising if one takes into account that
the nature of the 2D and 3D MHD turbulence is rather different.  In particular,
magnetic field wandering, which is an essential part of the LV99 reconnection
model, has a radically different nature in 3D and 2D.  What we have in 2D is
coherent displacement of magnetic field lines by Alfv\'enic perturbations, while
in 3D magnetic field lines can slip past one another,  enjoying the freedom provided by the
additional dimension.  As a result, in LV99 different bundles of wandering magnetic field
lines can enter the reconnection zone independently, and reconnect simultaneously.
It was shown in LV99 that it is this
simultaneous reconnection of independent magnetic flux bundles that makes
reconnection insensitive to resistivity.  This effect is absent in 2D where
the entry of fresh flux into the reconnection zone is constrained by the rate of
removal of reconnected flux.

The second effect that enables fast reconnection in the LV99 model is an
increase of the thickness of the outflow region.  There we see some similarity
between 2D and 3D reconnection.  Indeed, while the outflow in the Sweet-Parker
reconnection is constrained by the thin slot determined by Ohmic diffusivity, it
is due to an increase of the thickness of the outflow through turbulence that 2D
turbulent reconnection is faster than the Sweet-Parker rate.  The outflow
region in both 2D and 3D case depends on the intensity of driving and the scale
of turbulence injection.  However, the functional dependencies are different and
this difference stems from the difference in the reconnection physics.

As we mentioned earlier, magnetic field wandering plays a decisive role in
3D reconnection.  In 2D reconnection is not fast, since we observe the creation of
magnetic islands as a result of turbulence.  This process is rather limited
unless the turbulent injection velocity approaches $V_A$.  We believe that this
explains why the dependence of the reconnection rate on the Ohmic resistivity becomes
steeper as the turbulence weakens (see Figure~\ref{fig:vr_r}).  Indeed,  for weaker turbulence 
the alternate compression and expansion of the current sheet, which leads to magnetic island formation
is weaker and  reconnection is similar to  the 
Sweet-Parker model.  Similarly, the combination of anomalous resistivity and
magnetic island formation \footnote{While the 3D reconnection of weakly stochastic magnetic field
depends on the Alfv\'enic, i.e. the incompressible, component of magnetic
turbulence, the creation of magnetic islands depends on the compressible
component of the turbulence \cite[see][for the decomposition of MHD turbulent
motions]{cho-03}.} reaches a  saturated level of efficiency for modest values of the anomalous
resistivity, and appears to become insensitive to further increases  (see Figure~\ref{fig:vr_a}).

It is important to note that magnetic field structures, i.e. magnetic
islands, that we observe in 2D simulations do not appear in 3D simulation
\citep{kow-09}.  What is more, \cite{kow-09} obtained a stable value of
the reconnection speed in the presence of turbulence in a short period, about
one Alv\'en time.  In our work we are unable to reach this state because the
reconnection rate continuously and violently fluctuates.  Consequently 
the average speed of turbulent magnetic reconnection
in 2D is burdened with much higher statistical errors than in 3D.

In our work we determine the dependence of the reconnection rate on four
quantities: the power of turbulence $P_{inj}$, the injection wavenumber $k_f$, the
uniform $\eta_u$ and the anomalous $\eta_a$ resistivities.  The same analysis was
done by \cite{kow-09} in three dimensions.  For the power of turbulence and injection scale they
found the following scalings: $V_r^{TB}\sim P_{inj}^{1/2}$ and $V_r^{TB}\sim
l_{inj}^{3/4}$.  Compared to our findings the dependencies obtained in 3D are
stronger.  Namely, in the 2D case the reconnection rate grows with the power of
turbulence as $V_r^{TB}\sim P_{inj}^{1/3}$ and with the injection scale as
$V_r^{TB}\sim l_{inj}^{2/3}$.


\subsection{Earlier Studies of 2D Turbulent Reconnection and Their Relation to Work}

As we mentioned in the introduction, the concept of turbulent enhancement of
reconnection is not unprecedented.  In \cite{kow-09} we presented a list of papers
where turbulent effects were cited as the source of fast astrophysical
reconnection.  Compared to the present paper, all these papers lacked
a precise means of measuring the reconnection speed and therefore the numerical
simulations were providing mostly qualitative results.  Moreover, they 
adopted periodic
boundary conditions, which made it difficult to study turbulent reconnection
for more than a single Alfv\'en time.

Among these papers, the pioneering work were the numerical studies in
\cite{mat-85,mat-86}. There the analysis of 2D simulations of turbulence
revealed the formation of magnetic islands and X-points reminiscent of the
Petscheck process.

In our 2D model of turbulent reconnection we also observe continuous formation
of magnetic islands which are ejected from the reconnection zone due to open
boundary conditions.  For instance, we observe that when big loop of magnetic
field is removed from the box through boundaries, the reconnection rate
increases sharply.  After that the reconnection rate drops until new magnetic
island can be created.

The formation of islands is also a consequence of the tearing instability.  In
this vein, \cite{lou-04} examined the nonlinear growth of tearing modes and
obtained a fast growth of magnetic islands.  However, in their simulations they
used periodic boundary conditions, which keep such  islands in the reconnection
zone.

The most directly relevant work  is  \cite{fan-04,fan-05}.  They found that
the turbulent magnetic reconnection in solar atmosphere could be described by
a two phase process: first-slow and second-fast.  The rapid stage of reconnection
is caused by the coalescence instability \cite{bis-80,wu-01} which may also enhance
the reconnection rate in our simulations.

Here, and in contrast to the 3D model of  LV99,  we find that in 2D the influence of  the
uniform resistivity  on the reconnection
rate is stronger for higher values of $\eta_u$ and lower values of $P_{inj}$.
Similar results have also been obtained by \citep{lou-09}.  They claimed that
the reconnection process is fast in 2D only in some particular circumstances
i.e. for higher values of the Lundquist number and moderate levels of
turbulence.  We disagree with the claim of fast turbulent reconnection 
in 2D.  In fact, in the most of models analyzed by
\citep{lou-09} the reconnection process depends on the uniform resistivity.
Again an important  limitation of the aforementioned study is their limited averaging
arising from their choice of boundary conditions.

\subsection{Implications}

Two dimensional magnetic reconnection is the result of a rather artificial configuration.  The
value of our present study is that it clarifies the role of the effects of
dimensionality for the actual 3D reconnection of astrophysical magnetic fields
in the presence of weak turbulence.  The latter has many essential implications
starting from the First Order Fermi acceleration of energetic particles
\citep{gou-05,laz-09a} to Solar Flares \cite[see][]{laz-09b} and removal of
magnetic field during star formation \citep{lim-09}.  These
implications are discussed in more detail in \cite{kow-09}.

\section{Conclusions}
\label{conclusion}
In this  paper we have examined the results of 2D simulations of reconnection
process in presence of subAlfvenic turbulence. Our findings may be summarized as follows:

\begin{itemize}
\item Unlike Sweet-Parker reconnection, 
the reconnection of a weakly stochastic magnetic field is fundamentally different in 2D
and 3D,  in agreement the LV99 study. Reconnection in 2D
depends on resistivity and is not fast.

\item The enhancement of magnetic reconnection in 2D arises from an increase in the
thickness of the  plasma outflow layer due to the  creation of magnetic
islands which are ejected from the reconnection layer.

\item The power of turbulence $P_{inj}$ and the injection scale $k_f$ have an
influence on the reconnection rate. In our study this is:
$V_r^{TB}\sim P_{inj}^{1/3}\sim l_{inj}^{2/3}$.
\end{itemize}

\acknowledgements
This work of KK and KO was supported by the Polish Ministry of Science and
Higher Education through grants: 92/N-ASTROSIM/2008/0 and 3033/B/H03/2008/35.
GK's work was supported by the Polish Ministry of Science and
Higher Education through 3033/B/H03/2008/35 and by the National Science Foundation
through AST 0808118. The work of AL was supported by AST 0808118 and ATM 0648699. 
The work by EV was supported National Science and Engineering Research Council of Canada. 
The  computations presented here have been performed on the GALERA supercomputer in 
TASK Academic Computer Centre in Gda\'nsk.



\begin{thebibliography}{}
 \bibitem[Alvelius (1999)]{alv-99}
  Alvelius, K. 1999, Physics of Fluids, 11, 1880
 \bibitem[Armstrong et al.(1995)]{arm-95}
  Armstrong, J.~W., Rickett, B.~J., \& Spangler, S.~R. 1995, \apj, 443, 209
 \bibitem[Biskamp \& Welter(1980)]{bis-80}
  Biskamp, D., \& Welter, H. 1986, \prl, 44, 1069
 \bibitem[Biskamp(1996)]{bis-96}
  Biskamp, D. 1996, \apss, 242, 165
 \bibitem[Blackman(2000)]{bla-00}
  Blackman, E.~G., \& Field, G.~B. 2000, \apj, 534, 984
 \bibitem[Blackman \& Field(2008)]{bla-08}
  Blackman, E.~G., \& Field, G.~B. 2008, \mnras, 386, 1481
 \bibitem[Cattaneo \& Vainshtein (1991)]{cat-91}
  Cattaneo, F., Vainshtein, S.~I. 1991, \apj, 376, L21
 \bibitem[Cho \& Lazarian(2003)]{cho-03}
  Cho, J., \&  Lazarian, A. 2003, \mnras, 345, 325
 \bibitem[Del Zanna et al.(2003)]{del-03}
  Del Zanna, L., Bucciantini, N. \& Londrillo, P. 2003, \aap, 400, 397
 \bibitem[Drake et al.(2006a)]{dra-06a}
  Drake, J.~F., Swisdak, M., Che, H., \& Shay, M.~A. 2006a, \nat, 443, 553
 \bibitem[Elmegreen \& Scalo(2004)]{elm-04}
  Elmegreen, B.~G., \& Scalo, J. 2004, \araa, 42, 211
 \bibitem[Fan et al.(2004)]{fan-04}
  Fan, Q.-L., Feng, X.-S. \& Xiang, C.-Q. 2004, Physics of Plasmas, 12, 052901
 \bibitem[Fan et al.(2005)]{fan-05}
  Fan, Q.-L., Feng, X.-S. \& Xiang, C.-Q. 2005, Physics of Plasmas, 11, 12
 \bibitem[de Gouveia dal Pino \& Lazarian(2005)]{gou-05}
  de Gouveia dal Pino, E., \& Lazarian, A. 2005, \aap, 441, 845
 \bibitem[Gruzinov \& Diamond(1994)]{gru-94}
  Gruzinov, A.~V., \& Diamond, P.~H. 1994, \prl, 72, 1651
 \bibitem[Hanasz et al.(2004)]{han-04}
  Hanasz, M., Kowal, G., Otmianowska-Mazur, K., \& Lesch, H. 2004, \apjl, 605, L33
 \bibitem[Horbury \& Balogh(2001)]{hor-01}
  Horbury, T.~S. \& Balogh, A. 2001, \jgr, 106, 15929
 \bibitem[Innes et al.(1997)]{inn-97}
  Innes, D.~E., Inhester, B., Axford, W.~I., \& Wilhelm, K. 1997, \nat, 386, 811
 \bibitem[Kowal et al.(2009)]{kow-09}
  Kowal, G., Lazarian, A., Vishniac, E.~T. \& Otmianowska-Mazur, K. 2009, \apj, 700, 63
 \bibitem[Krause \& Radler(1980)]{kra-80}
  Krause, F. \& Radler, K.~H. 1980, {\em  Mean-Field Magnetohydrodynamics and Dynamo Theory}(Oxford: Pergamon Press)
 \bibitem[Lazarian \& Vishniac(1999)]{laz-99}
  Lazarian, A., \& Vishniac, E.~T., 1999, \apj, 517, 700, (LV99)
 \bibitem[Lazarian \& Vishniac(2000)]{laz-00}
  Lazarian, A., \& Vishniac, E.~T., 2000, Rev. Mex. Astron. Astrofis., 9, 55
 \bibitem[Lazarian \& Vishniac(2008)]{laz-08}
  Lazarian, A., \& Vishniac, E.~T. 2008, in Revista Mexicana de Astronom\'\i{}a y Astrof\'\i{}sica Conf. Ser., in press, (arXiv:0812.2019)
 \bibitem[Lazarian \& Opher(2009)]{laz-09a}
  Lazarian, A., \& Opher, M. 2009, \apj, in press, (arXiv:0905.1120)
 \bibitem[Lazarian et al.(2009)]{laz-09b}
  Lazarian, A., Vishniac, E., \& Kowal, G. 2009, Astronomical Society of the Pacific Conference Series, 406, 23
 \bibitem[Lesch(1993)]{les-93}
  Lesch, H., 1993, Conf. Proc. IAU Symposium 157, {\em  The Cosmic Dynamo} , ed. F. Krause, 395
 \bibitem[Lima et al.(2009)]{lim-09}
  Lima, R., Lazarian, A., \& de Gouveia dal Pino, E. 2009, in preparation
 \bibitem[Londrillo \& Del Zanna(2000)]{lon-00}
  Londrillo, P., \& Del Zanna, L. 2000, \apj, 530, 508
 \bibitem[Loureiro et al.(2004)]{lou-04}
  Loureiro, L., Cowley, S. C., Dorland, W. D., Haines, M. G. \& Schekochihin, P., 2009, astro-ph/0407.047
 \bibitem[Loureiro et al.(2009)]{lou-09}
  Loureiro, L., Uzdensky, N., Schekochihin, P., Cowley, S.C., \& Yousef, T.A., 2009, astro-ph/0904.0823
 \bibitem[Matthaeus et al.(1984)]{mat-84}
  Matthaeus, W.~H., Ambrosiano, J.~J., \& Goldstein, M.~L. 1984, \prl, 53, 1449
 \bibitem[Matthaeus \& Lamkin(1985)]{mat-85}
  Matthaeus, W.~H., \& Lamkin, S.~L. 1985, Physics of Fluids, 28, 303
 \bibitem[Matthaeus \& Lamkin(1986)]{mat-86}
  Matthaeus, W.~H., \& Lamkin, S.~L. 1986, Physics of Fluids, 29, 2513
 \bibitem[McKee \& Ostriker(2007)]{mck-07}
  McKee, C.~F., \& Ostriker, E.~C. 2007, \araa, 45, 565
 \bibitem[Mignone(2007)]{mig-07}
  Mignone, A. 2007, Journal of Computational Physics, 225, 1427
 \bibitem[Moffat(1978)]{mof-78}
  Moffat, H.~K. 1978, {\em Magnetic Field Generation in Electrically conducting Fluids}, (London, UK/New York, NY:Cambridge University Press)
 \bibitem[Pagano et al.(2008)]{pag-08}
  Pagano, P., Raymond, J.~C., Reale, F., \& Orlando, S. 2008, \aap, 481, 835
 \bibitem[Parker(1957)]{par-57}
  Parker, E.~N., 1957, \jgr, 62, 509
 \bibitem[Parker(1979)]{par-79}
  Parker, E.~N. 1979, {\em Cosmical Magnetic Fields}, (Oxford, USA: Clarendon Press)
 \bibitem[Parker(1992)]{par-92}
  Parker, E.N., 1992, ApJ, 401, 137
 \bibitem[Petschek(1964)]{pet-64}
  Petschek, H.~E. 1964, in Conf. Proc. of the AAS-NASA Symposium, {\em Physics of Solar Flares}, ed. W.~H. Hess, (Washington, DC:NASA Science and Technical Information Division), 425
 \bibitem[Priest \& Forbes(2000)]{pri-00}
  Priest, E., \& Forbes, T. 2000, {\em Magnetic Reconnection}, (Cambridge, UK:Cambridge University Press)
 \bibitem[Ruzmaikin et al.(1988)]{ruz-88}
  Ruzmaikin A. A., Shukurov A. M. \& Sokoloff D. D., 1988, {\em Magnetic Fields of Galaxies}, Ap\&SS Library, Kluwer Academic Publishers, Dordrecht
 \bibitem[Servidio et al.(2009)]{ser-09}
  Servidio, S., Matthaeus, W.~H., Shay, M.~H., Cassak, P.~A. \& Dmitruk, P., 2009, Phys. Rev. Lett., 101, 115003
 \bibitem[Sweet(1958)]{swe-58}
  Sweet, P.~A. 1958, Conf. Proc. IAU Symposium 6, {\em Electromagnetic Phenomena in Cosmical Physics}, ed. B. Lehnert, (Cambridge, UK:Cambridge University Press), 123
 \bibitem[T\'oth(2000)]{toth-00}
  T\'oth, G.,  2000, JCoPh, 161, 605
 \bibitem[Uzdensky(2006)]{uzd-06}
  Uzdensky, B. 2006, astro-ph/0607656
 \bibitem[Vainshtein \& Cattaneo (1992)]{vai-92}
  Vainshtein, S. I. \& Cattaneo, F., 1992, ApJ, 393, 165
 \bibitem[Vishniac \& Lazarian(1999)]{vis-99}
  Vishniac, E.~T., \& Lazarian, A. 1999, \apj, 511, 193
 \bibitem[Wu \& Chang(2001)]{wu-01}
  Wu, C.C., \& Chang, T. 2001, J. Atmos. Sol.-Terr. Phys., 63, 1447
 \bibitem[Yamada et al.(2006)]{yam-06}
  Yamada, M., Ren, Y., Ji, H., Breslau, J., Gerhardt, S., Kulsrud, R. \& Kuristyn, A., 2006, Physics of Plasmas, 13, 05, 2119
 \bibitem[Yokoyama \& Shibata(1995)]{yok-95}
  Yokoyama, T., \& Shibata, K. 1995, \nat, 375, 42
\end{thebibliography}
\end{document}